\documentclass[12pt,a4paper,english]{article}

\usepackage[bindingoffset=0.3cm,textheight=22.5cm,hdivide={2.7cm,*,2.7cm}, vdivide={*,22cm,*}]{geometry}
\usepackage[bookmarksnumbered=true,breaklinks=true]{hyperref}

\usepackage{amsmath,amsfonts,amssymb,babel,slashed}

\usepackage[numbers,square,comma,sort&compress]{natbib}

\IfFileExists{dsfont.sty}
	{\usepackage{dsfont}
         \newcommand{\id}{\mathds{1}}}
	{\typeout{Package dsfont.sty was not found, using alternative macros.}
         \let\mathds=\mathbb
         \newcommand{\id}{\mbox{1 \kern-.59em {\rm l}}}}
\let\one=\id

\makeatletter

         \@addtoreset{equation}{section}
\makeatother

\newcommand{\nocontentsline}[3]{}
\newcommand{\tocless}[3]{\bgroup\let\addcontentsline=\nocontentsline#1{#2}#3\egroup}

\newcommand{\qed}{\nobreak \ifvmode \relax \else
      \ifdim\lastskip<1.5em \hskip-\lastskip
      \hskip1.5em plus0em minus0.5em \fi \nobreak
      \vrule height0.75em width0.5em depth0.25em\fi}

\newenvironment{definition}[1][Definition]{\begin{trivlist}
\item[\hskip \labelsep {\bfseries #1}]}{\end{trivlist}}

\newcommand{\be}{\begin{equation}}
\newcommand{\ee}{\end{equation}}
\newcommand{\eq}[1]{(\ref{#1})}

\def\nn{\nonumber}
\def\bea{\begin{eqnarray}}
\def\eea{\end{eqnarray}}
\def\obar{\overline}

\def\beqa{\begin{eqnarray}} 
\def\eeqa{\end{eqnarray}} 
\def\beq{\begin{equation}} 
\def\eeq{\end{equation}}

\def\Tr{{\rm Tr}}

\def\a{\alpha}          
\def\b{\beta}           
\def\c{\gamma}  
\def\d{\delta}

\def\g{\gamma} 

\def\k{\kappa}
\def\l{\lambda} \def\L{\Lambda} \def\la{\lambda}
\def\m{\mu}     \def\n{\nu}

\def\r{\rho}
\def\s{\sigma}  

\def\th{\theta}

\def\vp{\varphi}

\def\cA{{\cal A}}  \def\cC{{\cal C}}
 \def\cE{{\cal E}} 
\def\cG{{\cal G}} \def\cH{{\cal H}} 
 \def\cK{{\cal K}} \def\cL{{\cal L}}
\def\cM{{\cal M}} \def\cN{{\cal N}} \def\cO{{\cal O}}
 \def\cQ{{\cal Q}} 
  \def\cU{{\cal U}}

\newcommand{\und}[1]{\underline{#1}}

\newcommand{\R}{\mathds{R}}
\newcommand{\C}{\mathds{C}}

\newcommand{\Z}{\mathds{Z}}

\newcommand{\msu}{\mathfrak{s}\mathfrak{u}}
\newcommand{\mso}{\mathfrak{s}\mathfrak{o}}
\newcommand{\mmu}{\mathfrak{u}}

\def\bit{\begin{itemize}}
\def\eit{\end{itemize}}

\def\({\left(}
\def\){\right)}

\def\Mat{{\rm Mat}}

\def\d{\delta}

\def\pa{\partial} \def\del{\partial}

\newcommand{\tr}{\mbox{tr}}

\def\bcomment#1{}

\def\LNC{\Lambda_{\rm NC}}
\def\YM{{\rm YM}}

\newcommand{\co}[2]{[#1,#2]}						
\renewcommand{\a}{\alpha}
\renewcommand{\b}{\beta}
\renewcommand{\d}{\delta}

\renewcommand{\th}{\theta}

\renewcommand{\l}{\lambda}

\renewcommand{\r}{\rho}

\newcommand{\G}{\Gamma}

\renewcommand{\L}{\Lambda}

\newcommand{\eps}{\varepsilon}
\newcommand{\Conf}{\mathcal{C}}
\newcommand{\Alg}{\mathcal{A}}
\newcommand{\B}{\mathcal{B}}
\newcommand{\Spin}{\mathrm{Spin}}
\newcommand{\rhs}{r.h.s.\ }

\newcommand{\wrt}{w.r.t.\ }
\newcommand{\cf}{cf.\ }
\newcommand{\const}{\mathrm{const}}

\newcommand{\NO}[1]{\colon \negthickspace #1 \! \colon \negthickspace }

\DeclareMathOperator{\diag}{diag}

 \allowdisplaybreaks[3]

\begin{document}

\renewcommand{\title}[1]{\vspace{10mm}\noindent{\Large{\bf
#1}}\vspace{8mm}} \newcommand{\authors}[1]{\noindent{\large
#1}\vspace{5mm}} \newcommand{\address}[1]{{\itshape #1\vspace{2mm}}}


\begin{flushright}
UWThPh-2013-04\\
CCNY-HEP-13/2
\end{flushright}

\begin{center}

\title{ \Large   Brane compactifications and  4-dimensional geometry\\[1ex] 
 in the IKKT model}

\vskip 3mm

 \authors{Alexios P. Polychronakos${}^{* ,}${\footnote{alexios@sci.ccny.cuny.edu }}, 
 Harold Steinacker${}^{\dagger ,}${\footnote{harold.steinacker@univie.ac.at}}, 
 Jochen Zahn${}^{\dagger ,}${\footnote{jochen.zahn@univie.ac.at}}
 }
 
\vskip 3mm

 \address{ 
  ${}^*$ {\it Physics Department\\
 The  City College of the CUNY \\ 
 160 Convent Avenue, New York, NY 10031, USA } \\[3ex]

${}^{\dagger}$  {\it Faculty of Physics, University of Vienna\\
Boltzmanngasse 5, A-1090 Vienna, Austria  }  
  }

\vskip 1.4cm

\textbf{Abstract}

\end{center}

We study in detail certain brane solutions with compact extra dimensions $\cM^4 \times \cK$ in the IKKT matrix model, with
$\cK$ being a two-dimensional rotating torus embedded in $\R^6$. We focus
on the compactification moduli and the fluctuations of $\cK \subset \R^6$ and their 
physical significance. Mediated by the  Poisson tensor, they
contribute to the effective 4-dimensional metric on the brane, and thereby become gravitational degrees of freedom. 
We show that the zero modes corresponding to the global symmetries of the model lead to 
Ricci-flat 4-dimensional metric perturbations, wherever the energy-momentum tensor vanishes.
Their coupling to the energy momentum tensor depends on the extrinsic curvature of the brane.

\vskip 1.4cm

\newpage

\tableofcontents

\section{Introduction}\label{sec:background}

Matrix models of Yang-Mills type are very interesting candidates for a 
theory of fundamental interactions including gravity. 
In particular, the so-called 
IKKT or IIB matrix model \cite{Ishibashi:1996xs} is singled out by maximal supersymmetry, 
and thus has a good chance to provide a well-defined quantum theory. 
The basic observation is that these models admit noncommutative or quantized 
submanifold (``branes'') as solutions. This leads to a relation with string theory and 
supergravity, 
and the model has been proposed as a non-perturbative definition of string theory;
cf. \cite{Chepelev:1997ug,Kabat:1997sa,Kitazawa:1998dd,Aoki:1999vr} for some basic references.
Here we follow the idea that suitable  brane solutions could play the role of physical space-time.
Indeed, fluctuations around such solutions give rise to 
noncommutative gauge theory living on the brane,
governed by a universal effective metric. 
This dynamical metric absorbs the $U(1)$ degrees of 
freedom of the gauge theory  \cite{Steinacker:2010rh}, and plays the role of a gravitational metric.  
Such an ``emergent'' gravity scenario is supported by several observations including 
gauge transformations giving rise to symplectomorphisms, (tangential)  
would-be $U(1)$ modes leading to Ricci-flat vacuum perturbations \cite{Rivelles:2002ez},
and other related observations \cite{Yang:2006hj,Steinacker:2009mp,Steinacker:2012ra,Steinacker:2012ct}.
However, it remains to be shown that the full Einstein equations emerge in a suitable regime.

In order to model realistic physics, basic branes such as $\R^4 \subset \R^{10}$ are clearly too simple.
One way to introduce additional structure as required for particle physics
is to consider compactified extra dimensions. 
In this paper, we discuss some specific new solutions of the IKKT model with compactified extra 
dimensions $\cM^4 \times \cK \subset \R^{10}$. 
These solutions behave for low energies as flat 4-dimensional
spaces with Minkowski signature. The extra  
$\cK$ arises from a fuzzy torus $T^2_N$ embedded in the 6 transversal dimensions of the model, which is stabilized by 
angular momentum and (generically non-vanishing) flux. This generalizes solutions found previously 
for the IKKT model \cite{Steinacker:2011wb} as well as the BFSS model, e.g. 
\cite{Bak:2001kq}, \cite{Hoppe:1997gr}.

Besides elaborating structural aspects of the  solutions, we focus on the 
effective 4-dimensional metric which governs the lowest Kaluza-Klein (KK) modes on the brane, 
and plays the 
role of a gravitational metric. As pointed out in \cite{Steinacker:2012ra},
the moduli of the extra dimensions directly affect the effective 4-dimensional metric,
due to the noncommutative structure.
Our aim is to understand these metric contributions due to the extra dimensions, and to clarify the effective 
 gravitational dynamics resulting from the matrix model action.

As a consequence of the global $SO(9,1)$ symmetry of the matrix model, the embedding of the compact space 
$\cK \subset \R^6$ in the transversal directions leads to massless zero modes, 
which  are nothing but Goldstone bosons from the 4-dimensional point of view. These zero modes are 
expected to play a central role in the low-energy or long-distance  physics on the brane.
We therefore focus on the dynamics of these zero modes, and clarify their 
contribution to the 4-dimensional curvature perturbations. It turns out that they lead indeed to Ricci-flat 
metric perturbations at locations without matter, $T_{\mu\nu} = 0$, provided the compactification has 
non-vanishing flux. The latter condition is imposed in order to stabilize the radial modes.
However, due to this radial stabilization, matter acts as a source for the 4-dimensional
Ricci tensor only via derivative terms $\del_\l T_{\mu\nu}$, 
similar to the contributions from the would-be $U(1)$ gauge fields  
\cite{Rivelles:2002ez,Steinacker:2010rh}. 
This complements and contrasts the results in \cite{Steinacker:2012ra} for 
the case of massless radial modes, where
a non-derivative coupling to $T_{\mu\nu}$ and hence a non-vanishing Newton constant was found. 
That however entails mixing between radial and tangential degrees of freedom,
which obscured the analysis leading to inconclusive results. 
For the present backgrounds, we conclude that the dynamics of the geometry 
is compatible with the vacuum sector of gravity, however the appropriate coupling to 
matter requires a different mechanism which is not seen in the present analysis.  
Such a coupling might arise in various ways on branes with extrinsic curvature 
\cite{Steinacker:2009mp,Steinacker:2012ra,Steinacker:2012ct}, which will be 
pursued elsewhere.

It is also interesting to consider the same type of backgrounds from the point of view of 
4-dimensional non-commutative gauge theory. We point out that they correspond to certain time-dependent 
  solutions which are periodic rather than translation invariant.
In particular, analogous solutions should also exist 
for conventional $\cN=4$ super-Yang-Mills theory, realized by time-dependent non-trivial VEVs
of the 6 scalar fields. However, in the absence of noncommutativity the $U(1)$ sector would decouple, and
the effective 4-dimensional geometry would not be affected 
by the compactification (i.e. the scalar fields). On the other hand, 
a similar effect is expected to arise on branes with flux embedded in $\R^{10}$ 
 governed by the Dirac-Born-Infeld action.

The approach to matrix models pursued here is rather different from much of the work 
in the literature.
The standard lore in string theory says that gravity originates from the closed string sector on
10-dimensional target space, which must subsequently be compactified to 4 dimensions. 
For an excellent  review including recent advances such as intersecting brane models 
see \cite{Blumenhagen:2006ci}.
Such a compactification of the target space leads to a vast landscape of vacua, with its inherent
lack of predictivity \cite{Susskind:2003kw}.
In the context of matrix models, analogous compactifications of the target space
were discussed in \cite{Connes:1997cr,Aoki:2010gv}, 
via a somewhat ad-hoc constraint on the matrices. 
In contrast, the present approach is based on the observation that the matrix model provides directly the 
world-volume description of branes $\cM \subset \R^{10}$, with effective (``open string'') 
metric captured by non-commutative gauge theory\footnote{Note that the 
matrix model is expected to be perturbatively finite on branes with 4 noncompact dimensions,
in contrast to the case of 10-dimensional compactifications \cite{Ishibashi:1996xs}.}. 
This suggests that a $4$-dimensional brane dynamics can arise without the need to compactify the target space.
If this effective 4-dimensional gravity turns out to be physically viable, 
the traditional 10-dimensional compactifications would no longer be needed.
This is the main motivation for the present approach.
Thus compactification here refers to brane solutions with structure $\cM^4 \times \cK \subset \R^{10}$, and
our aim is to study the dynamics of particular solutions of this type.
Some analogous solutions in string theory or supergravity are known, including
in particular the tubular brane solutions discussed in \cite{Bak:2001kq}, \cite{Emparan:2001ux}.
However, we are not aware of directly related results or works in the context of string or brane theory
which address their perturbations and dynamics.

This paper is organized as follows. After recalling some background we 
present the basic structure of the solutions under consideration in section \ref{sec:compact-split}, focusing on  
three classes of solutions characterized by non-vanishing currents. Their semi-classical significance is 
elaborated. We then explain the 4-dimensional gauge theory interpretation of the backgrounds 
in section \ref{sec:gaugetheory}. In section \ref{sec:perturb-zeromodes} we study the zero modes 
of the embedding fluctuations of $\cK$, 
elaborate their effective action, and determine the resulting perturbations of the Ricci tensor. 
Finally the appendices provide explicit details for the solutions under consideration
as well as a general discussion of conserved currents in the matrix model.

\section{Matrix models and their geometry} 
\label{sec:matrixmodels-intro}

We briefly collect the essential ingredients of the matrix model framework
and its effective geometry, referring 
to the recent review \cite{Steinacker:2010rh} for more details.

The starting point is given by a matrix model of Yang-Mills type, 
\begin{align}
S_{\YM} &= \frac {1}4\Tr\(\co{X^A}{X^B}\co{X^C}{X^D}\eta_{AC}\eta_{BD}\,  +  2 \obar\Psi_\a \tilde \gamma_A^{\a \b} [X^A,\Psi_\b] \)   
\label{S-YM}
\end{align}
where the $X^A, \Psi_\a$ are Hermitian matrices, i.e., operators acting on a separable Hilbert space $\mathcal{H}$. The index of $X$ runs from $0$ to $D-1$, and will be raised or lowered with  the invariant tensor $\eta_{AB}$ of $SO(D-1,1)$.
The index of $\Psi$ runs from $1$ to $2^{[D/2]}$, corresponding to the $D$-dimensional spinor representation. The matrices $\tilde \gamma_A$ are the corresponding $\gamma$ matrices.
Although this paper is mostly concerned with the bosonic sector, we focus 
 on the maximally supersymmetric IKKT or IIB model \cite{Ishibashi:1996xs} with $D=10$, which is best suited for quantization.
Then $\Psi$ is a matrix-valued Majorana Weyl spinor of $SO(9,1)$.
The model enjoys the fundamental gauge symmetry 
\begin{align}
X^A & \to U^{-1} X^A U, &  \Psi & \to U^{-1} \Psi U,
\label{gauge-inv}
\end{align}
where $U$ is a unitary operator on $\cH$, as well as the 10-dimensional Poincar\'e symmetry
\begin{align}
X^A & \to \L(g)^A_B X^b,  & \Psi_\a & \to \tilde \pi(g)_\a^\b \Psi_\b,  & g & \in \widetilde{SO}(9,1),  \nn \\
X^A & \to X^A + c^A \one, &  & & c^A & \in \R^{10},
\label{poincare-inv}
\end{align}
and a $\cN=2$ matrix supersymmetry \cite{Ishibashi:1996xs}.
The tilde indicates the corresponding spin group.
We  define the matrix Laplacian as
\begin{align}
  \Box \Phi :=  [X_B,[X^B,\Phi]]
\end{align}
for any matrix $\Phi \in \cL(\cH)$.
Then  the equations of motion of the model take the following form
\be
 \Box X^A \,=\, [X_B,[X^B,X^A]] =  0
\label{eom-IKKT}
\ee
for all $A$, assuming  $\Psi = 0$.

\subsection{Noncommutative branes and their geometry}

Now we focus on matrix configurations which describe embedded
noncommutative (NC) branes. This means that 
the $X^A$ can be interpreted as quantized embedding functions \cite{Steinacker:2010rh}
\be
X^A \sim x^A: \quad \mathcal{M}^{2n}\hookrightarrow \R^{10} 
\ee 
of a $2n$- dimensional submanifold of $\R^{10}$. More precisely,
there should be some quantization map $\cQ: \cC(\cM^{2n})  \to  \cA\subset \cL(\cH)$
which maps classical functions on $\cM$ to a noncommutative (matrix) algebra of functions, 
such that commutators can be interpreted as quantized Poisson brackets. 
In the semi-classical limit indicated by $\sim$,  matrices are identified with functions via $\cQ$,
in particular, $X^A = \cQ(x^A) \sim x^A$,
and commutators are replaced by Poisson brackets. For a more extensive introduction
see, e.g., \cite{Steinacker:2010rh}. 
Then the 
commutators
\begin{equation}
\co{X^A}{X^A} \sim i \{x^A, x^A\}(y) = i \th^{ab}(y)\del_a x^A(y) \del_b x^B(y)
\end{equation}
encode a quantized Poisson structure on $(\cM^{2n},\theta^{ab})$. Note that here and throughout, $x$ denote the embedding functions, and $y$ denote coordinates on $\cM^{2n}$.
This Poisson structure sets a typical scale of noncommutativity $\LNC$.
We will assume that $\th^{ab}$ is non-degenerate\footnote{If the Poisson structure is degenerate, 
then the fluctuations propagate only
along the symplectic leaves.}, so that the 
inverse matrix $\th^{-1}_{ab}$ defines a symplectic form 
on $\mathcal{M}^{2n}\subset\R^{10}$. This submanifold 
is equipped with the induced metric
\begin{align}
g_{ab}(y)=\pa_a x^A(y) \pa_b x_A(y),
\label{eq:def-induced-metric}
\end{align}
which is the pull-back of $\eta_{AB}$.  However, this is {\em not} the effective metric on $\cM^{2n}$. 
To understand the effective metric and gravity, we need to consider matter on the brane $\cM^{2n}$. 
Bosonic matter or fields arise from 
nonabelian fluctuations of the matrices around a stack $X^A \otimes \one_k$
of coinciding branes, while fermionic matter arises
from $\Psi$ in \eq{S-YM}.
It turns out that  in the semi-classical limit, the  effective action for such fields is
governed by a universal effective metric $G^{ab}$. 
It can be obtained most easily by considering the action of an additional 
scalar field $\phi$ coupled to the matrix model in a gauge-invariant way, with action
\begin{align}
 S[\phi] &= \frac{1}2\Tr [X_A,\phi][X^A,\phi] \nn\\
 & \sim -\frac{1}{2(2\pi)^n} \int d^{2n} y \sqrt{|\theta^{-1}|} \th^{aa'}\th^{bb'}g_{a'b'} \del_a\phi \del_b \phi \nn\\
 &= -\frac{1}{2(2\pi)^n}  \int d^{2n} y \sqrt{| G_{ab}|} G^{ab} \del_a\phi \del_b \phi .
 \label{action-scalar-geom}
\end{align}
Therefore  the effective metric is given by \cite{Steinacker:2008ri}
\begin{align}
 G^{ab}&= e^{-\sigma} \th^{aa'}\th^{bb'}g_{a'b'}\,,   \nn\\
  e^{-\s}&=\Big(\frac{\det{\th^{-1}}}{\det{ G_{ab}}}\Big)^{\frac 1{2}}
 = \Big(\frac{\det{\th^{-1}}}{\det{g_{ab}}}\Big)^{\frac 1{2(n-1)}}   
\label{eff-metric}
\end{align}
for $n>1$.
To understand the dynamics of the geometry in more detail, 
the following result is  useful \cite{Steinacker:2010rh}: the matrix Laplace operator 
reduces  in the semi-classical limit to the covariant Laplace 
operator\footnote{This result does not apply to the 2-dimensional case, where a modified 
formula holds \cite{arnlind-recent}.}
\begin{align}
 {\bf\Box} \Phi &= [X_A,[X^A,\Phi]]  
\ \sim\  -e^\s \Box_{G}\, \phi 
\end{align}
acting on scalar fields $\Phi \sim \phi$.
In particular, the matrix equations of motion \eq{eom-IKKT} take the  simple form
\be
0 =  \Box X^A \,\sim\, -e^\s \Box_{G} x^A .
\label{eom-IKKT-semi}
\ee
This means that the embedding functions $x^A \sim X^A$ are harmonic functions with respect to $G$.
Furthermore, the bosonic matrix model action \eq{S-YM} can be written in the semi-classical limit as follows
\begin{align}
 S_{\YM}\ \sim \ -\frac{1}{4(2\pi)^{2n}}\int d^{2n} y \sqrt{|\theta^{-1}|} \g^{ab}g_{ab} .
\label{action-semiclass}
\end{align}
Here we introduce
the  conformally equivalent  metric\footnote{More abstractly, this can be stated as 
$(\a,\b)_\g = (i_\a\theta,i_\b \theta)_g$ where $\theta = \frac 12\theta^{ab}\del_a\wedge\del_b$.}
\begin{align}
\g^{ab}&= \th^{aa'}\th^{bb'}g_{a'b'} = e^\s G^{ab}
\label{eff-metric-simple}
\end{align}
which satisfies
\begin{align}
\sqrt{|\theta^{-1}|} \g^{ab} &= \sqrt{|G_{ab}|} G^{ab}  .
\end{align}

\section{Compactified brane solutions and their geometry}
\label{sec:compact-split}

Now we focus on branes with compactified extra dimensions
\be
\cM^{2n}  = \cM^4 \times \cK \subset \R^D .
\label{compactification}
\ee
We will start with
explicit solutions where the extrinsic curvature is due to $\cK \subset \R^{10}$ 
while the embedding of $\cM^4$ is flat, and then proceed to study general perturbations 
around these solutions. 
That is, we consider embeddings
\be
 \cM^4 \times \cK \ni y \mapsto (x^\mu(y), x^i(y)) \in \R^4 \times \R^6 \cong \R^{10},
\ee
where $\mu \in \{ 0, \dots, 3 \}$, $i \in \{ 4, \dots, 9 \}$ in Cartesian coordinates on $\R^{10}$.
Such solutions including $\cK = T^2$ and $\cK = S^3 \times S^1$ 
have been given recently \cite{Steinacker:2011wb},
where $\cK$ is rotating along $\cM^4$ and stabilized by angular momentum. $\Box_G x^A = 0$ 
is possible because of  ``split" or mixed noncommutativity, where the
Poisson structure  relates the compact space $\cM^4$ with the non-compact space $\cK$,
\be
 \{x^\mu, x^i\} \neq 0 \ .
\label{split-NC}
\ee 
This implies that perturbations of $\cK$
lead to perturbations of the effective 4-dimensional metric on $\cM^4$, as elaborated below.

\subsection{The embedded fuzzy torus}
\label{sec:FuzzyTorus}

Starting with the unitary clock and shift matrices 
$U,V$ with $U V = q V U$ and $U^N = V^N = 1$ for $q= e^{2\pi i/N}$, we can define 3 complex 
or 6 hermitian matrices 
\begin{align}
{ Z^i} = \begin{pmatrix}
{ X^4 + i  X^5} \\
{ X^6 + i  X^7} \\
{ X^8 + i X^9}
\end{pmatrix}
\label{fuzzy-torus}
\end{align}
where
\begin{align}
  X^4 + i   X^5 & = U, &  X^6 + i  X^7 & = V, & X^8 & =  X^9 = 0 .
\label{torus-embedding}
\end{align}
This defines a fuzzy torus $T^2_N$ embedded in $\R^6$.
They satisfy the relations
\begin{align}
( X^4)^2 + ( X^5)^2 &= 1 = ( X^6)^2 + ( X^7)^2 ,  \nn\\
[ X^4,  X^5] &= 0 = [ X^6,  X^7]. 
\label{fuzzy-torus}
\end{align}
The irreducible representations of the  clock and shift matrices
are well-known \cite{sylvester} and need not be repeated here.
These matrices can be viewed as embedding maps
\be
 X^i \sim   x^i: \quad T^2 \hookrightarrow \R^{6} 
\ee 
and we can write 
\begin{align}
 X^4+i X^5 &= e^{i\Xi^4} \sim  x^4+i x^5 = e^{i\xi^4},   \nn\\
  X^6+i X^7 &= e^{i\Xi^5} \sim  x^6 + i x^7 = e^{i\xi^5}
\end{align}
with $\Xi^4, \Xi^5 \in \msu(N)$ and
\begin{equation}
\label{eq:T2_Poisson}
 \{\xi^4,\xi^5\} = \frac{\pi}{N}
\end{equation}
in the semi-classical limit.
The spectrum of the corresponding matrix Laplace operator is easily computed: 
\begin{align}
\Box \phi &= [ X^i,[ X^j,\phi]]\d_{ij} \\
\Box(U^n V^m) &= c ([n]_q^2 + [m]_q^2)\, U^n V^m 
\end{align}
where
\begin{align}
 [n]_q &=  \frac{\sin(n\pi/N)}{\sin(\pi/N)} \sim n  \nn\\
c &=   4 \sin^2(\pi/N)  
\end{align}
This implies 
\be
\Box X^i  
 =  4\sin^2(\pi/N)  X^i \
\qquad i \in \{ 4, \dots ,9 \}  .
\label{Box-torus}
\ee
Note that this relations holds trivially for $ X^8$ and $ X^9$.
We also note that the fuzzy torus enjoys a $\Z_N \times \Z_N$ symmetry implemented as gauge transformations
\begin{align}
 U  Z^i U^{-1} &= \begin{pmatrix}
                        1 & & \\  & q & \\ & & 1
                        \end{pmatrix}  Z^i, \qquad  
V  Z^i V^{-1} =  \begin{pmatrix}
                        q^{-1} & & \\  & 1 & \\ & & 1
                        \end{pmatrix}  Z^i \ .
\label{ZN-symm}
\end{align}
Finally, it should be obvious that the particular embedding chosen in \eq{torus-embedding}
can be generalized by acting with $SO(6)$ on the 6 matrices $X^a$. This will be exploited below.

To prepare the generalizations in the next section, it is useful to 
consider $i \cong \begin{pmatrix}
                       0 & 1 \\ -1 & 0
                      \end{pmatrix}$ as $\mmu(1)$ generator.
We can then identify the complex matrices $U,V$ with $2\times 2$ 
matrices with entries being hermitian $N\times N$ matrices,
via  
\begin{align}
U = e^{i \Xi^4} \cong \frac 12\begin{pmatrix}
                             U+U^\dagger & -i(U-U^\dagger)\\
                            -i(U^\dagger - U) & U + U^\dagger
                            \end{pmatrix}, \qquad
\label{U-complex}
\end{align}
and similarly for $V$.
With this identification,
the fuzzy torus embedding can be rewritten as
\begin{align}
{ Z^i} = \begin{pmatrix}
{ X^4 + i  X^5} \\
{ X^6 + i  X^7} \\
{ X^8 + i X^9}
\end{pmatrix}
= \diag_3(U, V, 0) {\bf z_0}
\label{complex-notation}
\end{align}
where ${\bf z_0} = (1, 1, 0) \in \C^3 \cong \R^6$, and $\diag_3$ indicates  a $3\times 3$
block-diagonal matrix with entries being $2\times 2$ matrices as above.

\subsection{Compactification with rotating fuzzy tori}
\label{sec:general}

In this section, we will exhibit an interesting class of solutions of the IKKT model 
with geometry $\cM^4 \times \cK$, where $\cK$ is a rotating version of the above torus.
The idea is to balance the brane tension with the centripetal force due to the rotation.
Some basic solutions of this type were given previously in \cite{Steinacker:2011wb} for the 
IKKT model, and in \cite{Bak:2001kq} for the BFSS model. We give a more general
setup which allows to study also their perturbations.

Let $\bar X^\mu \sim \bar x^\mu, \ \mu = 0,...,3$ generate the algebra of functions $\cA_\theta$
on the quantum plane $\R^4_\theta$, 
\begin{align}
 [\bar X^\mu,\bar X^\nu] = i \theta^{\mu\nu} ,
\label{Moyal-explicit}
\end{align}
with $\theta^{\mu\nu}$ in canonical block-diagonal form. 
Let $V_4 = e^{i \Xi^4}$, $V_5 = e^{i \Xi^5}$ be the generators of the fuzzy torus $T^2_N$, as
introduced in the previous subsection.
Now consider embeddings of the form
\begin{equation}
\label{eq:Ansatz}
 X^A = \begin{pmatrix} \bar X^\mu \otimes \one_N + \cE^\mu(\bar X, \Xi) \\ r(\bar X, \Xi) \cU_8 \dots \cU_4 {\bf z_0} \end{pmatrix}.
\end{equation}
where ${\bf z_0} \in \C^3$, $| {\bf z_0}| = 1$.
For $\a \in \{ 4, 5 \}$, we require
\begin{align}
 \cU_\a & = \cO \diag_3( U_\a^{n_\a^1}, U_\a^{n_\a^2}, U_\a^{n_\a^3}) \cO^{-1}, \nn \\
 U_\a & = e^{i \varphi^\a}, \nn \\
\label{eq:U_45}
 \varphi^\a & = k^\a_\mu \bar X^\mu + \Xi^\a + \cE^\a(\bar X, \Xi),
\end{align}
where $\cO \in SO(6)$, using a 
complex notation as in \eq{U-complex}, \eqref{complex-notation} for the last six embedding functions. 
For $\a \in \{ 6, 7, 8 \}$, we have
\begin{align}
 \cU_\a & = e^{\lambda_\a \varphi^\a}, \nn \\
 \varphi^\a & = k^\a_\mu \bar X^\mu + \cE^\a(\bar X, \Xi),
\label{eq:U_678}
\end{align}
with $\l_\a \in \mso(6)$. Introducing the notation
\begin{equation}
\label{eq:lambda_45}
 \l_\a = \cO \diag_3( i n_\a^1, i n_\a^2, i n_\a^3) \cO^{-1}
\end{equation}
for $\a \in \{ 4, 5 \}$, we may write
\begin{equation}
\label{eq:U_lambda}
 \cU_\a = e^{\l_\a \varphi^\a}
\end{equation}
for all $\a \in \{ 4, \dots, 8 \}$. The reason for requiring integer powers of $U_\a$, which leads to the form \eqref{eq:lambda_45}, is motivated by the desire to have a semiclassical interpretation in terms of continuous functions on the torus. This also leads to the requirement that the perturbations $\cE$ are polynomials in the torus generators, i.e., 
\begin{equation}
\label{eq:EpsExpansion}
 \cE^a(\bar X, \Xi) = \sum_{n} \cE^a_{n}(\bar X) e^{i (n_4 \Xi^4 + n_5 \Xi^5)}.
\end{equation}
The $\l_\a$ are required to be linearly independent, so that, with a supplementary condition discussed below, for constant radius $r$, all perturbations tangential to $S^5$ can be parametrized by the $\cE^\a$ (at least up to isolated points). This makes it possible to exploit the global $SO(6)$ symmetry. Further restrictions on the $\l_\a$ and $k^\a$ which ensure that the configuration \eqref{eq:Ansatz} describes a solution of the IKKT model will be discussed below.
The $\cE^a$ will be treated as perturbations, while $r$ will be assumed to be constant, as justified in Subsection~\ref{sec:FluxStabilization}.

In the following, we discuss such configurations in the semi-classical regime. In Appendix~\ref{sec:exact-solutions}, we show that in a certain limit, the semi-classical solutions we find correspond to solutions of the matrix model.

\subsection{Metric and semi-classical equations of motion}
\label{sec:eom}


Now we want to find sufficient conditions on the $\l_\a$ and $k^\a$ such that the ansatz \eqref{eq:Ansatz} with $\cE^a = 0$ and $r = \const$ gives solutions of the IKKT equations of motion in the semiclassical regime. Throughout, we work in Darboux coordinates $y^a = ( \bar x^\mu, \xi^i )$, where $\bar x^\mu$ and $\xi^i$ are the semiclassical counterparts of $\bar X^\mu$ and $\Xi^i$. Hence, the Poisson structure is given by
\begin{align}
 \theta^{ab} = \scriptsize{\begin{pmatrix}
                 \begin{pmatrix} 0 & \theta^{01} \\ -\theta^{01} & 0 \end{pmatrix}  & 0 & 0 \\
                  0 &  \begin{pmatrix} 0 & \theta^{23} \\ -\theta^{23} & 0 \end{pmatrix}  & 0 \\
                 0   & 0  & \begin{pmatrix}
                         0 &  \xi \\
                         - \xi & 0
                        \end{pmatrix}
               \end{pmatrix}},
\label{Poisson-full}
\end{align}
\cf \eqref{eq:T2_Poisson} and \eqref{Moyal-explicit}.
Henceforth, the notation $y^\mu$ indicates the restriction of the index $a$ to $\mu \in \{ 0, \dots, 3 \}$ and $y^i$ the restriction of $a$ to $i \in \{ 4, \dots, 5 \}$.
Then the semi-classical limit of the ansatz \eq{eq:Ansatz} takes the form
\begin{equation}
\label{eq:Ansatz-semiclass}
x^A = \begin{pmatrix}
       x^\mu \\  {\bf z}
      \end{pmatrix}
= \begin{pmatrix} y^\mu  + \cE^\mu(y) \\ r(y) \cU_8 \dots \cU_4 {\bf z_0}  \end{pmatrix},
\end{equation}
using the complex notation $\R^6 \ni \ x^i \ \cong {\bf z}  \ \in \C^3$.
Before discussing the equations of motions, let us first study the induced metric of these configurations 
in the semiclassical regime. The parametrization in \eqref{eq:Ansatz} is adapted to the embedding $\cK \subset S^5\subset \R^6$,
and the $\varphi^\a$, $\a \in \{ 4, \dots,8 \}$ can be viewed as local coordinates on $S^5$. 
The embedding metric on $S^5\subset \R^6$ in these local coordinates $\varphi^\a$ on $S^5$ is given by
\begin{equation}
 g^{(S^5)}_{\a\b} = \sum_{i=4}^9 (\l_\a x^i) (\l_\b x_i).
\label{gS5-def}
\end{equation}
In the following, we shall impose that $g^{(S^5)}$ has full rank, up to isolated points. This ensures that for the perturbations tangential to $S^5$ can be parametrized by the $\cE^\a$, or equivalently, that the $\varphi^\a$ are good coordinates on $S^5$.

If $\{ \la_4, \l_5 \}$ and $\{ \la_6, \l_7, \la_8 \}$ commute among themselves, as we will assume in the solutions of type A and B introduced below, this simplifies to
\begin{equation}
\label{eq:gS5_AB}
 g^{(S^5)}_{\a\b} =-r^2 {\bf z_0}^\dagger {\cU_5}^\dagger {\cU_4}^\dagger  \l_\a \l_\b  \cU_4\cU_5 {\bf z_0}.
\end{equation}
Note that under our assumptions, this is constant for $\a, \b \in \{ 4, 5 \}$, as the $\l$'s may be commuted past the $\cU$'s.
We can now compute the embedding metric and the effective metric on $\cM^6$ in Darboux coordinates:
\begin{align}
g_{ab} dy^a dy^b &= (\del_a x^\mu \del_b x^\nu \eta_{\mu\nu} +  \del_a x^i \del_b x_i) dy^a dy^b  \nn\\
&= \big(\eta_{\mu\nu} +  \del_{\nu} \cE^\rho\eta_{\mu\rho} + \del_{\mu}\cE^\eta\eta_{\eta\nu}  
 +  \del_\mu \cE_\eta \del_\nu \cE^\eta\big) dy^\mu dy^\nu + g_{ab}^{(\cK)}  dy^a dy^b \nn\\
\g^{ab} &= \theta^{aa'}\theta^{bb'} g_{a'b'} \ .
\end{align}
Here
\begin{align}
g_{ab}^{(\cK)} = \sum_{i=4}^9 \del_a x^i \del_b x_i \ &= \ \del_a\varphi^\a \del_b\varphi^\b g^{(S^5)}_{\a\b}  
 \ \stackrel{\cE = 0}{=} \  k_a^{\a}k_b^{\b}\, g^{(S^5)}_{\a\b} 
\label{g-int}
\end{align}
is the   contribution due to $\cK$. 
Using the notation $\bar g$ for the unperturbed value of $g$, i.e., for $\cE = 0$, we have
\begin{align}
 \bar g_{ab} &= \begin{pmatrix}
                \eta_{\mu\nu}  + g_{\mu\nu}^{(\cK)}  & g_{\mu j}^{(\cK)}  \\
                g_{i \nu}^{(\cK)}   & g_{ij}^{(\cK)} 
               \end{pmatrix}
             = \begin{pmatrix}
                \eta_{\mu\nu}  + k^\a_\mu k^\b_\nu  g_{\a\b}^{(S^5)}  & k^\a_\mu k^\b_j  g_{\a\b}^{(S^5)}  \\
                 k^\a_i k^\b_\nu g_{\a\b}^{(S^5)}  & k^\a_i k^\b_j  g_{\a\b}^{(S^5)}
               \end{pmatrix}.
\label{gbar-explicit}
\end{align}
Here we introduced a six-vector notation,
\begin{align}
 k^4 & = (k^4_\mu, 1, 0), & k^5 & = (k^5_\mu, 0, 1), & k^\a & = (k^\a_\mu, 0, 0), \quad \a \in \{ 6, 7, 8 \}.
\end{align}
The first order perturbation of $g_{ab}$ due to $\cE$ can easily be computed. For the component where $a, b$ are restricted to $\mu, \nu \in \{ 0, \dots, 3 \}$, we obtain
\be
 \delta_\cE g_{\mu \nu} = \del_{\nu} \cE^\rho\eta_{\mu\rho} + \del_{\mu}\cE^\eta\eta_{\eta\nu}  
 + \left( k^\a_\mu \del_\nu\cE^\b +\del_\mu\cE^\a k^\b_\nu \right) g_{\a\b}^{(S^5)}
  + k^\a_\mu k^\b_\nu \cE^\g \tfrac{\del}{\del \varphi^\g}   g_{\a\b}^{(S^5)}
\label{metric-perturb-egneral}
\ee
The crucial point is that linear fluctuations of the ''internal`` sector of the model contribute  
to the 4-dimensional metric, provided that $k_\mu^\a g_{\a\b}^{(S^5)} \neq 0$. 
This is a key difference from  ordinary (commutative) $\cN=4$ SYM theory, where 
the $U(1)$ sector completely decouples and the analogous internal perturbations would 
not contribute to the effective 4-dimensional metric.

Now we want to find sufficient conditions on the $\l_\a$ and $k^\a$ such that the ansatz \eqref{eq:Ansatz} with $\cE^a = 0$ and $r = \const$ gives solutions of the IKKT equations of motion in the semiclassical regime.
At the Poisson level, the equations of motion are
\begin{align}
 0 & = \{ x^A, \{ x_A, x^B \} \} \nn \\
    & = \theta^{aa'}\theta^{bb'}\del_a x^A \del_{a'} \left( \del_b x_A \del_{b'} x^B \right) \nn \\
    & = \theta^{aa'}\theta^{bb'} \del_{a'} \left( g_{ab} \del_{b'} x^B \right) \nn \\
    & = \del_{a} \left( \g^{ab} \del_{b} x^B \right) \nn \\
    & = (\g^{ab}  \del_{a} - e^\s\G^b)\del_{b} x^B, 
\end{align}
where we used that we are in Darboux coordinates,
and defined
\begin{equation}
\label{eq:def_Gamma}
 \G^b := -e^{-\s} \del_a \g^{ab}.
\end{equation}
For $x^B$ in the direction of $\cM^4$, i.e., $B = \mu \in \{0, \dots, 3 \}$, this reduces to
\be
 \Gamma^\mu = 0.
\ee
For the directions in which we embed $\cK$, we obtain
\begin{equation}
\label{eq:eom_AB}
 0 = r \cU_8  \cU_7  \cU_6 \left(-e^\s \Gamma^a k_a^{\a} \l_\a + \g^{ab} k_a^{\a} k_b^{\b} \NO{\l_\a \l_\b} \right) \cU_5 \cU_4 {\bf z_0}.
\end{equation}
Here we again assumed that $\{ \la_4, \l_5 \}$ and $\{ \la_6, \l_7, \la_8 \}$ commute among themselves. The colons indicate ordering in the order $\la_8, \l_7, \la_6, \l_5, \la_4$. Hence, a sufficient condition for a semiclassical solution is
\begin{align}
\label{eq:eom_1}
 \G^a & = 0, \\
\label{eq:eom_2}
 \g^{ab} k^\a_a k^\b_b \NO{\l_\a \l_\b} & = 0.
\end{align}
This will be discussed separately for three types of configurations.
We also note that \eq{eq:eom_1} together with \eqref{eq:gS5_AB} implies in particular 
\begin{align}
 \g^{ab}g_{ab}^{(\cK)} = 0
\label{radial-eq-first}
\end{align}
This determines the radius $r$, as discussed in more detail in Section~\ref{sec:FluxStabilization}.

\paragraph{Type A solutions.}

Type A solutions are characterized by $\cU_6 = \cU_7 = \cU_8 = \one$ 
(hence $k^{6} = k^{7} =k^{8} = 0$) and two commuting generators
$\l_\a$, $\a \in \{ 4,5 \}$ which satisfy
$-\l_4^2 + \l_5^2 =0$. These are supplemented by three commuting generators $\l_\a$, $\a \in \{ 6,7,8 \}$ such that $\{ \l_\a \}$ is linearly independent.
Then a sufficient condition to solve \eqref{eq:eom_2} is
\begin{align}
 \g^{ab} k_a^{\a} k_b^{\b} = p^2 \eta^{\a\b} = p^2\diag(-1,1)
\qquad \mbox{for}\ \ \a,\b \in \{ 4,5 \}
\label{k-orthog-A-general}
\end{align}
where $p \in \R$. 
As noted below \eqref{eq:gS5_AB}, $g^{(S^5)}_{\a \b}$ is constant for $\a, \b \in \{ 4,5 \}$,
so that both $g_{ab}$ and $\c^{ab}$ are constant, so in particular \eqref{eq:eom_1} is also satisfied.
Hence this class of solutions is characterized by two momenta 
$(k^{4},k^{5})$ which form an orthonormal 2-bein 
with respect to the effective  metric $\g^{ab}$.
Notice that this is in general a quartic equation in the $k_a^{\a}$ since
\begin{align}
 \g^{ab} k_a^{\a} k_b^{\b} = \eta_{\mu' \nu'} \theta^{\mu\mu'}\theta^{\nu\nu'} k_\mu^{\a} k_\nu^{\b}
  + \Theta^{\a\a'} \Theta^{\b\b'} g_{\a'\b'}^{(S^5)} 
\label{orth-full}
\end{align}
where 
\begin{align}
 \Theta^{\a\b} = \{ \varphi^{\a}, \varphi^{\b}\} = k^{\a}_ak^{\b}_b\theta^{ab} 
\label{theta-large-def}
\end{align}
We will see that type A solutions are characterized by 2 non-trivial constant currents
and a $\Z_N \times \Z_N$ symmetry.
The induced metric on $\cM^6$ in $(x^\mu,\varphi^\a)$ coordinates is constant
given by $(\eta_{\mu\nu}, g_{\a\b})$, and the structure is very similar to a quantum plane. In Appendix~\ref{app:Explicit_A}, we give explicit solutions of this type.

\paragraph{Type B solutions.}

Type B solutions are characterized by $\cU_7 = \cU_8 = \one $ 
 (i.e. $k^{7} =k^{8} = 0$) and three generators
$\l_\a,\ \a = 4,5,6$ which satisfy
$-\l_4^2 + \l_5^2 + \l_6^2=0$. These are supplemented by two generators $\l_7$, $\l_8$ such that $\{ \l_\a \}$ are linearly independent, and $\{ \l_4, \l_5 \}$ and $\{ \l_6, \l_7, \l_8 \}$ commute among each other. The momenta $k^\a$, $\a \in \{ 4, 5, 6 \}$ are chosen such that
\begin{equation}
\g^{ab} k_a^{\a} k_b^{\b} = p^2 \eta^{\a\b} = p^2 \diag(-1,1,1)
\qquad \mbox{for}\ \ \a,\b \in \{ 4,5,6 \} \label{k-orthog-B}.
\end{equation}
Hence, these solutions are characterized by three momenta 
$(k^{4},k^{5},k^{6})$ which form an orthonormal 3-bein
with respect to the effective  metric $\gamma^{ab}$, where $k^{4}$  is time-like.
Note that $\l_6$ does not commute with $\l_{4}, \l_5$, so that the metric $g_{\a \b}^{(S^5)}$ for $\a, \b \in \{ 4, 5, 6 \}$ will in general not be constant, \cf \eqref{eq:gS5_AB}. 
An explicit choice of $\la$'s and the corresponding metric $g_{\a \b}^{(S^5)}$ can be found in Appendix~\ref{app:Explicit_B}.
Using this explicit form \eq{kappa-explicit-B} of $g^{(S^5)}_{\a\b}$,
the condition \eqref{eq:eom_1}, i.e., $\del_a\g^{ab} = 0$, gives
\begin{align}
\theta^{aa'}\del_{a'} g_{ab}^{(\cK)}
&= \theta^{aa'}\del_{a'} \left( k_a^{5}k_b^{6} g^{(S^5)}_{56}(\varphi^4) + k_a^{6}k_b^{5} g^{(S^5)}_{65}(\varphi^4) \right) \nn\\
&= \left( \Theta^{46} k_b^{5} + \Theta^{45} k_b^{6} \right) \frac{\del g^{(S^5)}_{56}(\varphi^4)}{\del\varphi^4} = 0  
\label{del-g-vanish}
\end{align}
which requires $\Theta^{46} k_b^{5} =- \Theta^{45} k_b^{6}$, and therefore $\Theta^{46} = 0 = \Theta^{45}$.
Furthermore, there are potentially non-constant contributions in the orthogonality condition \eqref{k-orthog-B}
arising from $(\Theta^{\a 5} \Theta^{\b 6} + \Theta^{\a 6} \Theta^{\b 5})g^{(S^5)}_{56}(\varphi^4)$, \cf \eqref{orth-full}. This
implies that either $\Theta^{\a 5} = 0$ or $\Theta^{\a 6} = 0$. Therefore
\begin{align}
\Theta^{\a\b} = 0 \ , \qquad \a,\b = 4,5,6 \ ,
\label{theta-vanish}
\end{align}
which means that the $\varphi^{\a} = k^{\a}_a x^a$ form 3 mutually Poisson-commuting fields on $\cM^6$.
This also means that the quartic term in \eq{orth-full} drops out, and 
there is a non-empty moduli space of type B solutions as shown in Appendix~\ref{app:Explicit_B}.

\paragraph{Type C solutions.}

Type C solutions are characterized by $\cU_7 = \cU_8 = \one$ 
(i.e. $k^{7} = k^{8} = 0$)
and three mutually commuting generators
$\l_\a,\ \a = 4,5,6$ which satisfy
$-\l_4^2 + \l_5^2 + \l_6^2=0$. These are supplemented by two commuting generators $\l_7, \l_8$, such that $\{ \l_\a \}$ is linearly independent. This entails some obvious modifications in \eqref{eq:gS5_AB} and \eqref{eq:eom_AB}. One then chooses momenta $k^\a$, $\a \in \{ 4, 5, 6 \}$ such that
\begin{align}
 \g^{ab} k_a^{\a} k_b^{\b} = p^2 \eta^{\a\b} = p^2\, \diag(-1,1,1)
\qquad \mbox{for}\ \ \a \in \{ 4,5,6 \}.  \label{k-orthog-C-general}
\end{align} 
In contrast to type B, we do not need to impose  $\Theta^{\a\b} = 0$ since, by construction, $g_{\a \b}$ is independent of $\vp$ for $\a, \b = 4,5,6$, so that, as for type A,  
  $k^\a_a k^\b_b g_{\a\b}= \const$, and therefore  $\g^{ab}= \const$.
Hence these solutions are intrinsically flat, and turn out to support 3 constant currents 
 and  a $\Z_N \times \Z_N$ symmetry.
An example of a type C solution is given in Appendix~\ref{app:Explicit_C}.

In summary, we have obtained 3 types of compactified brane solutions of the semi-classical equations of motion.
For small $\Theta$, they correspond to exact matrix solutions of $\Box X^B =0$,
as shown  Appendix~\ref{sec:exact-solutions}.

\paragraph{Effective metric.}

The 4-dimensional effective metric which governs the long-distance physics on $\cM^4$, obtained by restricting $\gamma^{ab}$ to $a, b = \mu, \nu \in \{ 0, \dots, 3 \}$ in the Darboux coordinates \eqref{Poisson-full},
is given by
\begin{align}
 \g^{\mu\nu} &= \theta^{\mu \mu'}\theta^{\nu \nu'}\eta_{\mu'\nu'}  
+ \theta^{\mu a}\theta^{\nu b} k^{\a}_a k^{\b}_b  g_{\a\b},
\label{eff-metric-4D-1}
\end{align}
which is determined by the $k^{\a}$. 
The important point is that these $k_a^\a \approx \del_a \varphi^\a$ 
play a role similar to a vielbein,
and they are dynamical (albeit not independent as in general relativity).
Some of them will be related to conserved currents below.
This effective metric is constant for type A and C, and oscillating for type $B$ according to $g_{\a\b}^{(S^5)}$
(and should therefore be averaged at low energies).
Our aim is to understand the response of this metric to matter, which means to understand the effective
gravitational dynamics on $\cM^4$.

\subsection{Currents and conservation laws}

\label{sec:Currents}

Consider the  $SO(6)$ currents \eq{current-general} 
\begin{align}
J_{\und\a a} &=   (\l_{\und\a})_{ji} x^i\del_a x^j , \qquad \l_{\und\a} \in \mso(6)
\label{currents-NC}
\end{align}
which arise from matrix model currents as discussed in appendix \ref{app:LorentzCurrent}.
Here and in the following we denote with $\l_{\und\a}$ an arbitrary generator of $\mso(6)$, 
while $\l_\a$ indicates the particular generators chosen for $\a \in \{ 4,..,8 \}$.
In the absence of matter, these currents satisfy the following  conservation law
\begin{align}
\nabla^a J_{\und\a a}  &= 0.
\end{align}
For the above solutions, some of these currents are non-vanishing. 
Using the complex notation introduced in \eqref{eq:Ansatz-semiclass}, they can be computed via
\begin{align}
J_{\und\a a} &= \tfrac 12 (\l_{\und\a} {\bf z})^\dagger \del_a {\bf z} 
  + \tfrac 12 \del_a {\bf z}^\dagger \l_{\und\a} {\bf z},
\end{align}
which is complicated for a general $\l_{\und\a}$, and most currents vary along $\cK$. 
However for $\a \in \{ (6,) 7,8 \}$ corresponding to transversal deformations ($\a=6$ being excluded for type C),  they
take a simple form  related to the metric,
\begin{align}
J_{\a a}  = -\frac 12\sum_{\b}\del_a\varphi^\b {\bf z}^\dagger
   \big(\l_\a\l_\b  + \l_\b \l_\a \big) {\bf z}  
 &= \sum_{\b}\del_a\varphi^\b g_{\a \b}^{(S^5)}  , \qquad \a \in \{ (6,) 7,8 \}  
\end{align}
which holds even including perturbations.
For type A and C, the unperturbed currents for $\a \in \{ 4, \dots, 8 \}$ can be written as 
\begin{align}
  \bar J_{\a a} = k_a^\b g_{\b\a}^{(S^5)}, \qquad \a \in \{ 4, \dots, 8 \},
\label{J-explicit-BG}
\end{align}
where again the bar denotes the unperturbed quantity.
Using the explicit results for $g_{\a\b}^{(S^5)}$ in appendix \ref{sec:app-explicit}, 
we note that type A solutions have 2 constant currents $\bar J_\a \neq 0$, $\a \in \{ 4, 5 \}$ and type C
solutions have 3 constant currents $\bar J_\a \neq 0$, $\a \in \{ 4,5,6 \}$.
We expect that the solutions are to some extent characterized as ''ground states`` for these given currents.
Finally, the contribution from $\cK$ to the unperturbed
embedding metric \eq{gbar-explicit} can be written in terms of the currents as 
\begin{align}
 \bar g_{ab}^{(\cK)}  & =  k^\a_a \bar J_{\a b} =   k^\a_b \bar J_{\a a}. 
\label{eq:g-current-unperturbed}
\end{align}

\paragraph{Current conservation.}

As a consistency check, let us verify conservation of the currents $J_{\a}$, $\a \in \{ 4, \dots, 8 \}$ 
for the unperturbed type A and C  solutions. We can write the conservation law as
\begin{align}
 \del_a (\bar \g^{ab} \bar J_{\a b})  &=  -e^{\bar \s} \bar \G^{b} \bar  J_{\a b} +  \bar \g^{ab} \bar K_{\a ab}
, \qquad \a \in \{ 4, \dots,8 \}
\label{current-cons-explicit}
\end{align}
where we used the definition \eqref{eq:def_Gamma} and
\begin{align}
2 \bar K_{\a ab}  &:= \del_a \bar J_{\a b} + \del_b \bar J_{\a a} =  k_b^{\b} k_a^{\g}
    \left( \frac{\del \bar g_{\a\b}^{(S^5)}}{\del\varphi^\g} + \frac{\del \bar g_{\a\g}^{(S^5)}}{\del\varphi^\b} \right)
\label{Kab-def}
\end{align}
is related to the extrinsic curvature of the brane.
Since $\bar \G^{b}=0$ as verified in Section~\ref{sec:eom}, this leads to
\begin{align}
 \bar \g^{ab} \bar K_{\a ab} = p^2 \eta^{\b\g}\frac{\del \bar g_{\b\a}^{(S^5)}}{\del\varphi^\g} = 0
\end{align}
which follows from the orthogonality conditions \eqref{k-orthog-A-general}, \eqref{k-orthog-C-general}, and 
the explicit $g_{\a\b}^{(S^5)}$ in Appendix~\ref{sec:app-explicit}.
In principle, these conservation laws should completely capture the equations of 
motion for perturbations with fixed radius. 
However, we will not pursue this any further here.

We will see that the presence of these symmetries leads to massless (Goldstone) modes, 
including  the perturbations $\cE^\a$. 
Moreover they couple linearly to the metric, which implies that these are some sort of gravitational modes.

\subsection{Flux stabilization}
\label{sec:FluxStabilization}

We want to understand the dynamics of the compactification radius  $r$.
Assuming an un-perturbed compactification of the above type, 
the semi-classical matrix model action is given by 
\begin{align}
 S_{\rm YM} & \sim -\int \sqrt{|\theta^{-1}|} \g^{ab}g_{ab}  = -\int \sqrt{|G_{ab}|} V(r)  \nn\\
 V(r)  &= e^{-\s} \left( \eta_{\mu'\nu'}\theta^{\mu'\mu}\theta^{\nu'\nu} (\eta_{\mu\nu} 
  + 2 g_{\mu\nu}^{(\cK)})
  + \Theta^{\a\a'} \Theta^{\b\b'}  g_{\a\b}^{(S^5)}  g_{\a'\b'}^{(S^5)}   \right) 
\end{align}
recalling \eqref{gbar-explicit} and $\Theta^{\a\b} = \{\varphi^\a,\varphi^\b\}$, \cf \eqref{theta-large-def}.
Since $g_{\a\b}^{(S^5)} \sim r^2$, this gives a quartic potential $V(r) = V_0+ a r^2 + b r^4$
in the compactification radius $r$, which 
we consider as variable here.
Now we have to distinguish two cases. First, assume $\Theta^{\a\b}\not\equiv 0$, so that  there
is some flux on $\cK$. Then the quartic term in $V(r)$ 
is positive, leading to an effective potential 
for $r$ with minimum at $\bar r$ determined by
\begin{align}
0  =   \eta_{\mu' \nu'} \theta^{\mu\mu'}\theta^{\nu\nu'} g_{\mu\nu}^{(\cK)} 
 +  \Theta^{\a\a'}\Theta^{\b\b'}  g_{\a\b}^{(S^5)}  g_{\a'\b'}^{(S^5)} 
= \g^{ab} g_{ab}^{(\cK)} \ .
\end{align}
This coincides with the condition \eq{radial-eq-first} found previously.
In order to have $\bar r^2>0$ we must have $\eta_{\mu' \nu'} \theta^{\mu\mu'}\theta^{\nu\nu'} g_{\mu \nu}^{(\cK)} < 0$, 
so that the potential has a unique minimum at $\bar r^2 >0$ and mass 
\begin{align}
 m^2_r = V''|_{\bar r} = - 4 \eta_{\mu' \nu'} \theta^{\mu\mu'}\theta^{\nu\nu'} g_{\mu\nu}^{(\cK)} \  > 0 \ .
\end{align}
The scale is set by $r$ and the noncommutative structure $\theta^{ab}$, which are both UV scales.
This means that the radial perturbations are stabilized by the flux and massive, and we can 
safely set $r = \const$ at low energies. 

On the other hand if $\Theta^{\a\b}\equiv 0$, then the potential $V(r)$ is flat, leading to a massless 
radial mode.  Although this mode is interesting because it couples to the 
energy-momentum tensor \cite{Steinacker:2012ra}, it probably acquires a mass via quantum corrections since it is not 
protected by any symmetry.  We therefore focus on the case $\Theta^{\a\b} \neq 0$ from now on, 
for type A or C solutions.

\section{Gauge theory interpretation}
\label{sec:gaugetheory}

The solutions found above were interpreted up to now 
in terms of a brane with 6-dimensional effective geometry.
Now we use the interpretation of the the matrix model as noncommutative $\cN=4$ SYM 
theory on $\R^4_\theta$ with gauge group $U(N)$, via
\begin{align}
 X^A = \begin{pmatrix}
        \bar X^\mu \otimes \one_N \\ Z^i
       \end{pmatrix} .
\end{align}
Here the transversal matrices are renamed as
\begin{align}
 Z^i = \begin{pmatrix}
        \phi^1+i\phi^2 \\
         \phi^3+i\phi^4 \\
         \phi^5+i\phi^6
       \end{pmatrix}
\end{align}
and interpreted as 6 scalar fields on $\R^4_\theta$ in the adjoint representation of $U(N)$. 
The $SO(9,1)$ symmetry of the model then decomposes into $SO(3,1) \times SO(6)$, where
$SO(6)$ is the R-symmetry group of $\cN=4$ SYM.
The emergence of fuzzy spaces in nonabelian gauge theory
is well-known by now, and the
equivalence of these two interpretations of the matrix model constitutes the starting point
underlying emergent gravity  matrix models \cite{Steinacker:2010rh}.
More specifically,  we interpret the solutions 
\eqref{eq:Ansatz} in terms of coherent plane wave excitations of the 6 
$U(N)$-valued scalar fields $Z^i$, propagating along 2 resp.\ 3
momenta $k^{\a}$. ''Coherence`` here refers to particular $\msu(N)$ structure which is chosen such that
an effective toroidal geometry arises for large $N$.

The field-theoretic view is useful here for at least two reasons.
First, it allows to compute and compare different 
solutions and their currents resp. energy-momentum tensors, and select the 
preferred (lowest-energy) solutions for a given set of quantum numbers.
Second, it makes manifest the UV finiteness of the model, since the VEV of the scalar fields
becomes irrelevant in the UV where the model reduces to the  $\cN=4$ model on $\R^4_\theta$.
Nevertheless, this interpretation does not alter the fact that the effective metric for 
excitations around these solutions is given by \eq{eff-metric}, so that perturbations lead to 
a modified effective 4-dimensional metric. This is the key difference compared with  commutative $\cN=4$
SYM theory.

\subsection{Translational invariance and periodicity}
\label{sec:periodic}

Without perturbations,
the ansatz \eqref{eq:Ansatz} defines a periodic structure on the non-compact space $\cM^4 = \R^4$
defined by the 2 or 3 non-vanishing momenta $k^{\a}_\mu$. 
We can introduce a reciprocal basis $a^\mu_{\a}$ for the subspace 
spanned by these momenta supplemented by vectors $b^\mu_{\a'}$ such that 
\begin{align}
 k_\mu^{\b} a^\mu_{\a} & = 2\pi\d^{\b}_{\a}, 
  & k_\mu^{\b} b^\mu_{\a'} & = 0.
\end{align}
Then clearly the $\phi^i$ are invariant under $\bar x^\mu \to \bar x^\mu + a^\mu_{\a}$,
and $b^\mu_{\a'} \del_\mu \phi^i(\bar x)=0$.
These translations can be implemented by gauge transformations on $\R^4_\theta$ via
\begin{align}
\phi^i(\bar X^\mu+a^\mu)= T_a \phi^i(\bar X^\mu) T_a^{-1} = \phi^i(\bar X^\mu), \qquad T_a = e^{i a^\mu\theta^{-1}_{\mu\nu} \bar X^\nu} 
\end{align}
for $a=a_{\a}^\mu$, and 
\begin{align}
\phi^i(\bar X^\mu+b^\mu) =  T_b \phi^i T_b^{-1} = \phi^i(\bar X^\mu)
\end{align}
for $b=b^\mu_{\a'}$.
Moreover, the lattice spanned by the $a^\mu_\a$, $\a \in \{ 4, 5 \}$,
has a  sub-structure defined by the $\Z_N\times \Z_N$ symmetry of the fuzzy tori, 
which amounts to a discrete translation invariance\footnote{The compactification of the IKKT model
considered in \cite{Connes:1997cr} are characterized by  similar relations, but were interpreted
in a very different way. The present considerations show that such solutions
 appear as non-compact periodic backgrounds for perturbations which propagate on them.}
\begin{align}
 V_4 \phi^i(\bar X^\mu) V_4^{-1} &=   \phi^i(\bar X^\mu +\tfrac 1N a^\mu_{4})
\  \stackrel{A,C}{=} \  \big(e^{\frac{2\pi}{N}\l_4}\big)^i_{\ j}\,  \phi^j(\bar X^\mu) , \nn\\
 V_5 \phi^i(\bar X^\mu) V_5^{-1} &=   \phi^i(\bar X^\mu+\tfrac 1N a^\mu_{5})
 \ \stackrel{A,C}{=} \ \big(e^{\frac{2\pi}{N}\l_5})^i_{\ j}\,  \phi^j(\bar X^\mu)
\label{A-C-symm-discrete}
\end{align}
The last equality holds only\footnote{The conjugation with $V_{4,5}$ induces a 
$\l_{4,5}$ factor between the operators $\cU_8 \cU_7 \cU_6$ and $\cU_5 \cU_4$ in \eqref{eq:Ansatz}, which for type A and C
can be commuted to the left since, for type A, $\cU_\a = \one$ for $\a \in \{ 6,7,8 \}$ and, for type C, $\cU_\a = \one$ for $\a \in \{ 7,8 \}$ and $\l_6$ commutes with $\l_{4,5}$.} for type A and C, 
which thereby respect a global $\Z_N \times \Z_N$ R-symmetry
up to gauge transformations.
Furthermore for type B and C  we have 
\begin{align}
 e^{2\pi c \l_6} \phi^i(\bar X) &=  T_{c a_{6}}^{-1} \phi^i(\bar X^\mu)\, T_{c a_{6}} 
\end{align}
for arbitrary $c$.
In the semi-classical limit $N \to \infty$, we can introduce the 6 generators
\begin{align}
P_\mu &= \theta^{-1}_{\mu \nu} \{X^{\nu},.\} = \del_\mu ,  \nn\\
P_i &=\theta_{ij}^{-1}\{\xi^j,.\} =  \del_i \ .
\label{P-l-relation}
\end{align}
Then the above discrete lattice symmetry implies
\begin{align}
 \left( a^\mu_{\a} P_\mu -  P_\a \right) \phi^i = 0  \ & \stackrel{A,C}{=} \  \left( \delta^i_j a^\mu_{\a} P_\mu - 2\pi {\l_\a}^i_{\ j} \right) \phi^j , \qquad \a \in \{ 4,5 \}    \nn\\
 0 \ & \stackrel{B,C}{=} \ \left( \delta^i_j a^\mu_{6} P_\mu -  2\pi{\l_6}^i_{\ j} \right) \phi^j.
\label{A-C-symm-cont}
\end{align}
Therefore the solutions under considerations are not vacua in the 
usual sense of quantum field theory, but can be considered as 
''generalized vacua`` which enjoy a  discrete translational invariance  analogous to solid state theory.
This  discrete translational invariance 
should characterize the states under consideration.
In view of the enhanced symmetry  \eq{A-C-symm-discrete} resp.\ \eq{A-C-symm-cont},
type A or C seem to  be more natural candidates for ''vacuum`` geometries.

\subsection{Kaluza-Klein modes}

For the toroidal compactifications under consideration, all modes
(both for the geometry as well as for matter or gauge fields) can be decomposed into 
Kaluza Klein (KK) modes,
\begin{align}
 \Phi(y) = \sum_{n,m} \Phi_{n,m}(y^\mu) e^{i n y^4 + m y^5}
  \equiv \sum_{n,m = -N/2}^{N/2} \Phi_{n,m}(y^\mu) V_4^n V_5^m.
\end{align}
Even though the metric $\g^{ab}$ does not respect the product structure $\cM^6 = \cM^4 \times \cK$, for type A and C $\gamma^{ab}$ is constant, i.e., the effective Laplacian respects the $U(1) \times U(1)$ symmetry (more precisely, the $\Z_N \times \Z_N$ symmetry),
and therefore the above decomposition.
Explicitly,
\begin{align}
 \Box \Phi = \sum_{n,m} V_4^n V_5^m \, \big( - c_{n,m} \Phi_{n,m} + 2 i A^\mu_{n,m} \del_\mu \Phi_{n,m}
  + \del_\mu(\g^{\mu\nu} \del_\nu \Phi_{n,m})\big)
\end{align}
with
\begin{align*}
 c_{n,m} & = \g^{44} n^2 + 2 \g^{45} n m + \g^{55} m^2, \\
 A^\mu_{n,m} & = \g^{\mu 4} n + \g^{\mu 5} m.
\end{align*}
Again, the $\mu, \nu$ indices only run in $\{0, \dots, 3 \}$. Setting
\[
 \nabla^{n,m}_\mu = \del_\mu + i \tilde \gamma_{\mu \nu} A^\mu_{n,m},
\]
where $\tilde \gamma_{\mu \nu}$ is the inverse of $\gamma^{\mu \nu}$, we may write the wave equation for $\Phi_{n,m}$ as
\[
 \left( \gamma^{\mu \nu} \nabla^{n,m}_\mu \nabla^{n,m}_\nu + A^\mu_{n,m} \tilde \gamma_{\mu \nu} A^\nu_{n,m} - c_{n,m} \right) \Phi_{n,m} = 0.
\]
The ``vector potentials'' $A^\mu_{n,m}$ simply shift the origin of momentum space. Regarding stability, it is thus important to check that $A^\mu_{n,m} \tilde \gamma_{\mu \nu} A^\nu_{n,m} - c_{n,m}$ is negative. We may write it as
\[
 A^\mu_{n,m} \tilde \gamma_{\mu \nu} A^\nu_{n,m} - c_{n,m} = \begin{pmatrix} n & m \end{pmatrix} Q \begin{pmatrix} n \\ m \end{pmatrix}
\]
with some $2 \times 2$ matrix $Q$. For our type A and C solutions one explicitly checks that this matrix is negative definite, so we indeed have stability of the KK modes.
 

The above reduction to 4-dimensions by keeping only the trivial KK modes in $V_{4,5}$
can be written more geometrically  as
\begin{align}
\langle \cdot \rangle := \frac 1{(2\pi)} \int \sqrt{ | \theta_{(T_2)}^{-1} |} d y^4 d y^5 \sim  \tr_N (\cdot).
\label{average-def}
\end{align}
Here $\theta_{(T_2)}$ is the restriction of $\theta^{ab}$ to $a,b \in \{ 4, 5\}$ in the Darboux coordinates \eqref{Poisson-full}.
For type A and C, this amounts to an averaging procedure
over the compactification $\cK$. However for type B, it  effectively averages also over a unit cell of 
the 4-dimensional periodicity identified in section \ref{sec:periodic}. 
We can then introduce the reduced effective metric $G^{\mu\nu}_{(4D)}$ 
\begin{align}
G^{\mu\nu}_{(4D)} &= e^{-\s_{(4D)}} \g^{\mu\nu}_{(4D)}, 
&  \g^{\mu\nu}_{(4D)} & = \langle  \g^{\mu\nu} \rangle , 
& e^{-\s_{(4D)}} & = \frac{\sqrt{|\theta_{\mu \nu}^{-1}|}}{\sqrt{| G^{(4D)}_{\mu\nu}|}},
\label{eff-metric-4D}
\end{align}
cf.\ section 5.1 in \cite{Steinacker:2012ra}. Here $G^{(4D)}_{\mu\nu}$ is the inverse of $G^{\mu\nu}_{(4D)}$ and $\theta_{\mu \nu}^{-1}$ and $\gamma^{\mu \nu}$ are the restriction of $\theta_{ab}^{-1}$ and $\g^{ab}$ to $a,b = \mu, \nu \in \{ 0, \dots, 3 \}$ in Darboux coordinates \eqref{Poisson-full}.
This metric governs the action for the lowest KK modes.
For example, the action for a scalar field $\phi(x)$ takes the form 
\begin{align}
 S[\phi] = \int d^4 y \sqrt{|\theta_{\l \r}^{-1}|} \g^{\mu\nu}_{(4D)} \del_\mu \phi \del_\nu \phi
 = \int d^4 y \sqrt{| G^{(4D)}_{\l \r}|}  G^{\mu\nu}_{(4D)} \del_\mu \phi \del_\nu \phi,
\end{align}
recalling that the Poisson structure separates nicely in Darboux coordinates.

\section{Geometry, perturbations and curvature}
\label{sec:perturb-zeromodes}

\subsection{Perturbations and coupling to matter}
\label{sec:PerturbationsMatter}

We are now restricting to type A and C and consider first order perturbations. For these types, we have 
\begin{align}
 k_a^\alpha k_b^\beta \tfrac{\del}{\del\varphi^\gamma}  g_{\alpha \beta}^{(S^5)} =0
 \label{AC-k-property}
\end{align}
as $g_{\a \b}^{(S^5)}$ is constant for $\a, \b \in \{ 4, 5, (6) \}$, \cf Section \ref{sec:eom}. It follows that the first order perturbation of the induced metric is given by
\be
\label{eq:delta_g}
 \delta_\cE g_{ab} = \begin{pmatrix} \del_{\nu} \cE^\rho \eta_{\mu\rho} + \del_{\mu}\cE^\eta\eta_{\eta\nu}  
 + \bar J_{\a \mu} \del_\nu\cE^\a +  \bar J_{\a \nu} \del_\mu\cE^\a & \bar J_{\a \mu} \del_j \cE^\a +  \bar J_{\a j} \del_\mu\cE^\a \\
\bar J_{\a i} \del_\nu \cE^\a + \bar J_{\a \nu} \del_i \cE^\a & \bar J_{\a i} \del_j \cE^\a + \bar J_{\a j} \del_i \cE^\a
   \end{pmatrix},
\ee
where we used \eqref{J-explicit-BG}.


Now we want to include matter to the system. Since matter couples to the effective metric
$G^{ab}$, the variation  of the matter Lagrangian with respect to the geometry is given as usual in terms of the 
energy-momentum tensor $T_{ab}$ of matter,
\begin{align}
 \delta S_{M} &= \int \sqrt{|\bar G_{ab}|} T_{ab} \delta G^{ab} 
 =  \int \sqrt{|\bar G_{ab}|} e^{-\bar \sigma} T_{ab} \Pi^{ab,cd} \d g_{cd}.
\label{e-m-coupling}
\end{align}
Here we note that the variation of the effective metric can be written as
\[
 \delta G^{ab} = e^{- \bar \sigma} \delta \gamma^{ab} + \delta e^{-\sigma} \bar \gamma^{ab} 
= e^{- \bar \sigma} (\theta \delta g \theta)^{ab} - \tfrac{1}{2(n-1)} \bar g^{cd} \delta g_{cd} \bar G^{ab} 
= e^{- \bar \sigma} \Pi^{ab,cd} \d g_{cd},
\]
where
\[
 \Pi^{ab, cd} = \theta^{ac} \theta^{db} - \tfrac{1}{2(n-1)} \bar g^{ab} \bar \gamma^{cd},
\]
and a bar stands for the unperturbed quantity.
Using \eqref{eq:delta_g}, this becomes 
\begin{align}
 \delta S_{M} &= -2\int \sqrt{|\theta^{-1}|} \left[ \cE^\a
\left( \bar K_{\a ab} \Pi^{ab,cd} T_{cd} + \bar J_{\alpha b} \Pi^{ab,cd} \del_a T_{cd} \right) + \cE^\mu \eta_{\mu \nu} \Pi^{a \nu, c d} \del_a T_{cd} \right] 
\label{e-m-coupling-PI}
\end{align}
using \eq{J-explicit-BG} and \eq{Kab-def}.
Therefore the presence of matter leads to perturbations of $\cE^\a$
mediated by $\bar J_\a \neq 0$. Non-derivative coupling to matter 
arises in the presence of extrinsic curvature $\bar K_{\a ab} \neq 0$.
This induces a dynamical rotation of $\cK\subset\R^6$ along $\cM^4$, which in turn affects the effective geometry.
This will be elaborated in more detail for the zero modes below.

\subsection{Zero modes and and low-energy effective action}
\label{sec:ZeroModes}

The matrix model action is invariant under the 10-dimensional Poincare group  $SO(9,1) \ltimes \R^{10}$.
This symmetry implies that given a solution, we get a new, degenerate solution\footnote{This global 
rotation preserves the type of solution (type A,B,C) under consideration, since
it simply rotates generators $\l_\a$.}  by acting with some group element. 
As usual, this leads to massless Goldstone bosons, and 
it is plausible that these zero modes govern the low-energy 
or long distance physics of the perturbed solutions.  
We therefore study these zero modes and their geometrical significance in detail.
For a related discussion focusing on the particle physics aspects see \cite{Nishimura:2012rs}.

Consider first the $SO(6)$ symmetry 
\begin{align}
\d_{\und\a} x^i = (\l_{\und\a})^i_{\ j} x^j\, , \qquad \l_{\und\a} \in \mso(6)
\end{align}
which acts on $\cK\subset \R^6$ and preserves the  non-compact brane $\cM^4 = \R^4$.
The corresponding Goldstone bosons are obtained by making these transformations $y^\mu$- dependent, 
\begin{align}
\d_{\und\a} x^i(y) = \L^{\und\a}(y^\mu) (\l_{\und\a})^i_{\ j} x^j(y).
\label{so6-mode}
\end{align}
They all describe different deformations of $\cK \subset S^5 \subset \R^6$ with fixed radius.
Therefore, this gives $\dim(\mso(6)) = 15$ Goldstone bosons on $\R^4$, some of which may be trivial for backgrounds with remaining symmetries (such as our example of an unperturbed type A background, which is invariant under rotations in the $8-9$ plane).
Along with the remaining 4-dimensional zero modes due to the other symmetries, 
they should govern the low-energy physics.
This is nothing but the usual low-energy effective field theory approach.

One of our assumptions in Section~\ref{sec:eom} was that the phases $\varphi^\a$ are coordinates of $S^5$, at least up to isolated points. Hence, an infinitesimal rotation of the tori can be written as\footnote{Note that the coordinates $\varphi^\a$ are only defined up to isolated points. However, in the arguments given below, this does not pose any problem, as these do not rely on the existence of coordinates, but only on the symmetry.}
\begin{equation}
(\l_{\und\a})^i_{\ j} x^j 
 = x^i(\varphi^\a + \cE^\a_{{\und\a}}) - x^i(\varphi)
 = \sum_{\a = 4}^8 \cE_{\und \a}^\a \ \frac{\del x^i}{\del \varphi^\a} 
\label{infinites-rotation}
\end{equation}
with infinitesimal $\cE^\a_{\und\a}(\varphi)$. Although we will not need the $\cE^\a_{{\und\a}}$ explicitly, 
it means that some of these zero modes correspond to non-trivial Kaluza-Klein modes on $\cK$. 
Hence the apparently new degrees of freedom $\L_{\und\a}(y^\mu)$ simply capture 
certain higher KK modes of $\cE^\a$, 
corresponding to $y^\mu$--dependent symmetry transformations of the  rigid objects $\cK$.
In a more complete treatment, 
we should expand the most general perturbation $\cE^\a(y)$ into  harmonics on $\cK$, 
obtain the equations of motion for these KK modes, 
and discard those who acquire a mass from the 4-dimensional point of view. 
This is a  non-trivial but well-defined task, which requires to solve the general equations of motion.
The approach followed below is based on symmetries and allows to 
short-cut this complex procedure in a simple and intuitive way.

There is an interesting alternative point of view. If these zero modes 
describe all relevant low-energy modes, then the low-energy effective action 
can be viewed as an action for a group-valued field on $\cM^4 = \R^4$, 
\begin{align}
 R(y^\mu) = \exp(\L^{\und\a}(y^\mu)\l_{\und \a}): \quad \cM^4 \to \cG \ .
\label{group-exp}
\end{align}
This is the case if the action of $\cG$ on the solution is free, which should hold
quite generically for sufficiently complex compactifications. 
Otherwise, one has to replace $\cG \to \cG/\cG_s$ where $\cG_s$ is the stabilizer group.

In order to proceed, it is advantageous to choose a basis  of $\mso(6)$ as follows,
\begin{align}
 \{ \tilde \l_{\und\a}\} = \{ \tilde \l_4, \tilde \l_5, \tilde \l_6; \tilde \l_{\und\a'} \},
\end{align}
such that the $\tilde \l_4, \tilde \l_5, \tilde \l_6$ are (in a suitable basis) mutually commuting
$2\times 2$-block-diagonal matrices\ of $\mso(6)$, 
 orthogonal (w.r.t the Killing metric) to the remaining 
 block-off-diagonal matrices $\tilde \l_{\und\a'}$. For type C, we simply choose $\tilde \l_\a = \l_\a$ for $\a \in \{ 4, 5, 6 \}$, and complements with $\tilde \l_{\und \a}$. For type A, we set $\tilde \l_\a = \l_\a$ for $\a \in \{ 4, 5 \}$, and choose $\tilde \lambda_6$ such that $\l_6 z_0 = 0$. A suitable choice for a $\tilde \l_6$ in our example for type A is given in Appendix~\ref{app:Explicit_A}.


\paragraph{Averaged currents.}

The unperturbed $\mso(6)$ currents \eqref{currents-NC} can be written as follows
\begin{align} 
 \bar J_{\und\a a} &= \sum_{\b = 4}^{5 (6)} H_{\und\a \b} \del_a \varphi^\b,  \nn\\
  H_{{\und\a} {\und\b}} &:= - \vec x \tilde \l_{\und\a} \tilde \l_{\und\b} \vec x,  
 \label{kappa-eps-2}
\end{align}
where $\vec x = x^i$ and $6$ is included in the sum for type C. Note that $H_{\a\b} = g_{\a\b}^{(S^5)}$ for $\a,\b \in \{ 4, 5, (6) \}$, where $6$ is included for type C, \cf \eq{gS5-def}. 
The average over $\cK$ can be written in the form
\begin{align}
  \langle H_{\und\a \und\b}\rangle = -\tr_N \left( \Pi \tilde \l_{\und\a} \tilde \l_{\und\b} \right)
\label{H-average}
\end{align}
where $\Pi$ is defined in  \eq{Pi-def}.
We verify explicitly in appendix \ref{sec:app-explicit} that $\Pi$ commutes with the three 
commuting $U(1)$ generators $\tilde \l_{4,5,6}$. 
It follows that $\langle H_{\und\a \b}\rangle$ is invariant under these  $U(1)$ subgroups
for $\b \in \{4,5,6 \}$, and therefore
\begin{align}
 \langle H_{\und\a \b}\rangle &=    0,    \quad \b \in \{4,5,6 \}, \ \ \und \a\not\in \{4,5,6\}   \nn\\[1ex]
 \langle \bar J_{\und\a a} \rangle &= \left\{\begin{array}{ll}       
                                         k^\b_a g_{\b\a}^{(S^5)}, &  \und \a \equiv \a \in \{4,5, (6)\}\\
                                         0,    &  \und \a\not\in \{4,5, (6)\} \ 
                                        \end{array} \right.    
\label{J-average}
\end{align}
where $\a=6$ is included for type C. To see this, note that in the unperturbed case $\del_a \varphi^\b$ is constant. Also note the condition $\tilde \l_6 z_0 = 0$ for type A.

To proceed, we will expand the action to quadratic order in the $\L^{\und\a}(y^\mu)$,
and study the associated perturbations of the 4-dimensional geometry.

\paragraph{Metric perturbations due to zero modes.}

We determine the metric perturbations due to the above zero modes.
Consider first the above $SO(6)$ modes.
From now on, we drop the tilde on $\tilde \l_\a$.
Similar as in \eqref{eq:delta_g}, we have
\begin{align}
 \d_\L g_{ab}^{(\cK)} &= \del_a  \vec x\del_b (\L^{\und\a} \l_{\und\a}  \vec x)
  + \del_a(-  \vec x\L^{\und\a} \l_{\und\a}) \del_b  \vec x \nn\\
   &=  \del_b \L^{\und \a} \bar J_{\und\a a} + \del_a \L^{\und \a} \bar J_{ \und \a b},
\label{eq:delta_Lambda_g}
\end{align}
which vanishes for constant $\L^{\und\a}$ as it should.
Here $\bar J_{\und \a a}$ is the $SO(6)$ current \eq{currents-NC}. 
Note that the sum is now over all the $\mso(6)$ generators rather than just 
$\a = 4,...,8$, which is sometimes useful \cite{Steinacker:2012ct}.
Similarly,  the translational symmetries
\begin{align}
 X^A \ \to X^A + c^A
\end{align}
 give rise to 10 Goldstone bosons $c^A(y^\mu)$.
The corresponding metric perturbations are
\begin{align}
 \d_{c} g_{ab} &= \del_a  c_A\del_b  x^A  + \del_a x^A \del_b  c_A  
  =  \del_a  c_A L_b^A  + L_a^A \del_b  c_A
\label{eq:delta_c_g}
\end{align}
where
\begin{align}
 L_b^A := \del_b x^A.
\end{align}
However, the 
zero modes corresponding to translations in the direction $\R^6$ in which $\cK$ is embedded do not couple to matter since $\langle L^A\rangle = 0$ for $A \in \{ 4, \dots, 9 \}$.

Formally, analogous considerations apply 
to the full $SO(9,1)$ symmetry. However, 
they are not expected to lead to independent physical modes, since the currents
 corresponding to the breaking modes of $SO(9,1) \to SO(3,1)\times SO(6)$
always vanish upon averaging  over $\cK$; moreover they diverge at infinity 
(cf. \cite{Nishimura:2012rs}). 
Similarly, the $SO(3,1)$ modes are redundant with the translational 
modes $c^\mu$. This leaves only the $SO(6) \times \R^6$ modes discussed above.
Nevertheless, it might be useful to keep track of the full $SO(9,1)$ symmetry if the embedding of 
the non-compact brane $\cM^4 \subset \R^{10}$ is non-trivial, as expected, e.g., for cosmological solutions 
 \cite{Klammer:2009ku}. 
This  will be pursued elsewhere.

\paragraph{Radial mode.}

For a configuration with $\Psi = 0$ for which the action \eqref{S-YM} vanishes, the scaling $X^A \to \a X^A$ is also a symmetry, with associated zero mode $\L^{(R)}$.
The corresponding metric perturbation is 
\begin{align}
 \d_R g_{ab}^{(\cK)} &= 2 \L^{(R)} g_{ab} + \frac 12 \del_b \L^{(R)} \del_a r^2 + \frac 12 \del_a \L^{(R)} \del_b r^2 
\end{align}
This is interesting because it provides a non-derivative coupling to the energy-momentum tensor.
Similarly, there might also be a symmetry $X^i \to \a X^i$ if $\Theta^{\a\b} = 0$.
However it seems likely that these radial modes are massive, 
and do not contribute to the low-energy physics.
In particular, this happens in the presence of flux on $\cK$ as explained in section \ref{sec:FluxStabilization}.

\paragraph{Coupling to matter.}

The coupling of these zero modes to matter is obtained  from \eq{e-m-coupling-PI},
\begin{align}
 \delta S_{M} &= -2\int_{\cM^4} \sqrt{|\theta^{-1}_{\l \r}|}  \left[ \L^{\und \a}
\left( \langle \bar K_{\a \mu\nu}\rangle T_{\mu'\nu'} 
+ \langle \bar J_{\und \a \nu} \rangle  \del_\mu T_{\mu'\nu'} \right) + c^\r \eta_{\r \nu}   \del_\mu T_{\mu'\nu'} \right] \Pi^{\mu\nu,\mu'\nu'}.
\label{e-m-coupling-average}
\end{align}
We assume here that matter responsible for the energy-momentum tensor is in the lowest KK mode, 
so that the energy-momentum tensor consists only of lowest KK modes and does not have any components along $\cK$.

\paragraph{Second order expansion and effective action.}

To get the action expanded up to second order in the zero modes, we need 
\begin{align}
\d^2_{\L}  \vec x &= \tfrac 12 \L^{\und\a} \L^{\und\b} \l_{\und \a} \l_{\und \b} \vec x  \nn\\
 \d^2_{\L} g_{ab}^{(\cK)} &= \del_a \left( -\vec x \L^{\und\a} \l_{\und\a} \right) 
  \del_b \left( \L^{\und\b} \l_{\und\b} \vec x \right) + \tfrac 12 \del_a  \vec x \del_b \left( \L^{\und\a} \L^{\und\b} \l_{\und \a} \l_{\und \b} \vec x \right)
 + \tfrac 12 \del_a \left( \vec x\L^{\und\a} \L^{\und\b} \l_{\und \a} \l_{\und \b} \right) \del_b  \vec x  \nn\\
 &= - \tfrac 12 f_{\und \a\und \b}^{\ \ \und\g} 
 \left( \bar J_{\und\g a}\del_b \L^{\und\a} \L^{\und\b} + \bar J_{\und \g b}\del_a \L^{\und\a} \L^{\und\b} \right)
   + \del_a \L^{\und\a}\del_b\L^{\und\b}\ H_{\und\a\und\b}  \nn
\end{align}
where again $\vec x = x^i$ and $f_{\und \a\und \b}^{\ \ \und\g}$ are the structure constants of $\mso(6)$.
The mixed variations are
\begin{align}
\d^2_{\L c} x^i &= \L^{\und\a}(\l_{\und\a})^i_j c^j   \nn\\
 \d^2_{\L c} g_{ab}^{(\cK)} &= \del_a c^i  \del_b \big(\L^{\und\a}(\l_{\und\a})_{ij}  x^j\big)  
  - \del_a\big(\L^{\und\a} x^i (\l_{\und\a})_{ij}\big)  \del_b c^j 
 + \del_a x^i \del_b\big(\L^{\und\a}(\l_{\und\a})_{ij} c^j \big)
 - \del_a\big(\L^{\und\a} c^i(\l_{\und\a})_{ij} \big)\del_b  x^j  \nn\\
 &=  \del_b \L^{\und\a} \big( \del_a c^i (\l_{\und\a})_{ij}  x^j + \del_a x^i (\l_{\und\a})_{ij} c^j \big)
  + (a \leftrightarrow b) \ .  \nn
\end{align}
Therefore the effective action for the zero modes expanded to second order is
\begin{align}
 S_{YM} &= \int \sqrt{\theta^{-1}} \theta^{aa'}\theta^{bb'}
  \left( \d g_{ab}\d g_{a'b'} + 2 g_{ab}\d^2 g_{a'b'} \right)  \nn\\
 &=  \int \sqrt{\theta^{-1}} 
  \Big( 2\theta^{aa'}\theta^{bb'} \left( \del_b \L^{\und\a} \bar J_{\und \a a} + \del_a \L^{\und \a} \bar J_{\und \a b} 
    + \del_a c_A L_b^A + \del_b c_A L_a^A \right)
  \left( \del_{b'} \L^{\und \b} \bar J_{\und \b a'} + \del_{b'} c_B L_{a'}^B \right) \nn\\
 & \quad - 2 \g^{ab} f_{\und \a\und \b}^{\ \ \und\g} \bar J_{\und \g a}\del_b \L^{\und \a} \L^{\und\b} 
  + 2 \g^{ab} \del_a \L^{\und \a}\del_b \L^{\und \b}  H_{\und \a \und \b}   \nn\\
 & \quad  + 4 \g^{ab} \del_b \L^{\und \a} \left( \del_a c^i (\l_{\und\a})_{ij}  x^j + \del_a x^i (\l_{\und\a})_{ij} c^j \right)
 + 2 \g^{ab} \del_a c^A \del_b c_A \Big).
\end{align}
Now recall that the 4-dimensional Goldstone bosons $\L^{\a}$ and  $c^A$ are constant along $\cK$.
Therefore we can write this action using the averaging $\langle .\rangle$ over $\cK$ introduced 
in \eq{average-def}. This simplifies using partial integration 
using $\langle  \bar J_{\und\a b}  \bar J_{\und\b a}\rangle= \const$ and $\langle x^i \rangle =0$, and
\begin{align}
  S_{YM} &= 2 \int d^4y \sqrt{ |\theta_{\l \r}^{-1}|}  \left( -
   \langle f \, f \rangle 
  - \langle \g^{a\n} \rangle f_{\und \a \und \b}^{\ \ \und\g}  \langle \bar J_{\und \g a} \rangle \del_\n \L^{\und\a} \L^{\und\b} 
  +  \g^{\m \n}_{(4D)} \del_\m \L^{\und\a}\del_\n \L^{\und\b}   \langle  H_{\und \a\und \b}\rangle 
  + \g^{\m \n}_{(4D)} \del_\m c^A \del_\n c_A \right) 
\end{align}
where 
\begin{align}
 f(\L,c) = \theta^{\mu\nu} \left( \del_\mu\L^{\und\a} \bar J_{\und\a \nu} + \del_\mu c^A  L_{A \nu} \right) .
 \label{gauge-fixing-function}
\end{align}
We note that 
$\langle  H_{\und \a\und \b} \rangle$ is non-degenerate for type C
due to \eq{H-average} and \eq{Pi-def}, and is invariant 
under the 3 commuting $U(1)$ generators. 
Pretending that all these modes are independent\footnote{Apart from the pure gauge modes, which could
be fixed by setting $\langle f \rangle = 0$. We assume here that the 
modes are constant along $\cK$. If the modes are not independent then there might be additional solutions.}, 
we obtain the equations of motion
\begin{align}
 \langle  H_{\und \a\und \b}\rangle e^{\s_{(4D)}} \Box_{G_{(4D)}} \L^{\und\b} - \theta^{\mu\nu} \del_{\mu}\langle \bar J_{\und\a \nu} f \rangle
 - \langle \g^{a \nu} \rangle f_{\und \a \und \b}^{\ \ \und\g} \langle \bar J_{\und\g a}\rangle \del_\nu \L^{\und\b}\
& = \langle \bar J_{\und\a \mu}\rangle \del_\nu \tilde T^{\mu\nu}   \nn\\
e^{\s_{(4D)}} \Box_{G_{(4D)}} c^A  - \theta^{\mu\nu} \del_{\mu}\langle L_{\nu}^A f \rangle 
  &= \langle L^A_\mu \rangle  \del_\nu \tilde T^{\mu\nu} 
\label{eom-zeromodes}
\end{align}
where
\begin{align}
\tilde T^{\mu\nu} &= \theta^{\mu\mu'} \theta^{\nu\nu'} T_{\mu' \nu'}  
\end{align}
A similar structure was obtained in \cite{Steinacker:2012ra}.
It follows from the explicit form \ \eqref{J-average} of $\langle \bar J_{\und\a}\rangle$ 
and $\langle H_{\und\a\und\b}\rangle$ that matter  $T_{\mu\nu}$ induces
perturbations only for 
the  $\L^{4,5,(6)}$, $6$ being included for type C, and the tangential translation modes $c^\mu$. In vacuum,
 the zero modes will be shown to imply Ricci-flat  perturbations.
This is perfectly consistent with gravity in vacuum,
however the appropriate coupling to matter must arise in a different way.
Some possible mechanisms will briefly be discussed below.

It should be clear that the results of this section are not restricted to the specific compactifications
under considerations but apply more generally.

\subsection{Linearized curvature tensor}

In this section we compute the linearized Ricci or Einstein tensor due to the above zero modes, 
for the effective 4-dimensional metric $G^{\mu\nu}_{(4D)}$, \cf \eqref{eff-metric-4D}. We only consider the case of type A or type C unperturbed background,
which is intrinsically flat.
Throughout, a bar will indicate the background value, obtained by setting $\cE^a = 0$, \cf Section~\ref{sec:general}.
We will work in Darboux coordinates, which are in metric compatible
with the background for type A and C, so that $\bar\nabla \equiv \del$.
The linearized perturbation of the effective 4-dimensional metric is given by
\begin{align}
\label{eq:h_munu}
h^{\mu \nu} & := \d G^{\mu \nu}_{(4D)} = e^{- \bar \s_{(4D)}} \d \g^{\mu \nu}_{(4D)} - \bar G^{\mu \nu}_{(4D)} \d \s_{(4D)},   & 
\d \s_{(4D)}  & =  - \tfrac 1{2} \bar G_{\mu \nu}^{(4D)} h^{\mu \nu} ,
\end{align}
using \eqref{eff-metric-4D}.
The linearized Ricci tensor for a perturbation $h^{\mu \nu}$ on a flat background $G^{\mu \nu}$ is given by \cite[Section~4.4]{Wald:1984rg}
\begin{align}
 \d R^{\mu \nu} &=  \tfrac 12 \Box_{ G} h^{\mu \nu}  +\tfrac 12 \del^\mu \d\G^\nu +  \tfrac 12 \del^\nu \d\G^\mu, \nn\\
 \d \G^\mu & = - \del_\nu h^{\mu \nu} + \tfrac{1}{2} \del^\mu ( G_{\lambda \rho} h^{\lambda \rho} ) = - \d \left( \tfrac 1{\sqrt{|G|}} \del_\nu \left( \sqrt{|G|}G^{\mu \nu} \right) \right).
\label{lin-Ricci-general} 
\end{align} 
In the present case, we have
\[
 \d \G^\mu_{(4D)} = - \delta \left( e^{- \sigma_{(4D)}} \del_\nu \g^{\mu \nu}_{(4D)} \right),
\]
\cf \eqref{eff-metric-4D}. Therefore, using \eqref{eq:h_munu}, the linearized Einstein tensor is 
\begin{align}
 \d \cG^{\mu \nu} &=  \d R^{\mu \nu} - \tfrac 12 \bar G^{\mu \nu}_{(4D)} \d R  \nn\\ 
 &= \tfrac 12 e^{-\bar \s_{(4D)}} \Box_{\bar G_{(4D)}} \d \g^{\mu \nu}_{(4D)} +\tfrac 12 \del^\mu \d \G^\nu_{(4D)} + \tfrac 12 \del^\nu \d \G^\mu_{(4D)} 
  - \tfrac 12 \bar G^{\mu \nu}_{(4D)} \del_\l \d\G^\l_{(4D)} 
\label{lin-Einstein-general} 
\end{align} 

Our aim is now to compute this for perturbations given by the zero modes identified in the previous subsection.
For type A and C, we get, with \eqref{eq:delta_Lambda_g}, \eqref{eq:delta_c_g}, and using \eqref{gauge-fixing-function},
\begin{align}
 (\d_{\L} +  \d_{c}) \g^{\m \n}_{(4D)} 
 &= \theta^{\m \l} \theta^{\n \r} \left( \del_{\r} \L^{\und \a} \langle \bar J_{\und \a \l} \rangle 
  + \del_{\r} c^A \langle L_{A \l} \rangle  \right) + (\m \leftrightarrow \n), \\
 e^{\bar \s_{(4D)}} (\d_\L + \d_c) \Gamma^\mu_{(4D)} &=  -
 \theta^{\mu \l} \theta^{\nu \r} \del_\nu \left( \del_{\r} \L^{\und \a} \langle \bar J_{\und \a \l}\rangle 
 + \del_{\l} \L^{\und \a} \langle \bar J_{\und\a \r} \rangle
+ \del_{\r} c^A \langle L_{A \l} \rangle + \del_{\l} c^A \langle L_{A \r} \rangle  \right)  \nn\\
 &= - \theta^{\mu \l} \del_{\l} \langle  f\rangle,
\label{eq:deltaGamma}
\end{align}
since $\langle \bar  J_{\und\a b}\rangle, \langle L_{A b}\rangle = \const$ and $G^{\mu \nu}_{(4D)}$ is constant, so that the first order variation of $\Gamma^\mu_{(4D)}$ is only sensitive to the variation of $\gamma^{\mu \nu}_{(4D)}$.
Now we use the equations of motion \eq{eom-zeromodes} and \eq{kappa-eps-2}, which gives 
\begin{align}
\Box_{\bar G_{(4D)}} \d_{\L} \g^{\mu \nu}_{(4D)} &=  \theta^{\mu \l} \theta^{\nu \r}
 \del_{\r} \Box_{\bar G_{(4D)}} \L^{\und \a} \langle H_{\und \a \b} \rangle k^\b_{\l}  + (\mu \leftrightarrow \nu) \nn\\
 &= e^{- \bar \s_{(4D)}} \theta^{\mu \l} \theta^{\nu \r} \Big( k^\b_{\l} \theta^{\sigma \xi} \del_{\r} \del_{\sigma} \langle \bar J_{\b \xi} f \rangle
  + f_{\a \und \b}^{\ \ \und \g} \langle \g^{a \sigma} \rangle k^\a_{\l} \langle \bar J_{\und \g a} \rangle \del_{\r} \del_\sigma \L^{\und \b} \nn\\
 & \qquad \qquad \qquad \qquad \qquad \qquad \qquad \qquad \qquad \qquad +  \langle \bar J_{\a \sigma} \rangle k^\a_{\l}  \del_{\r} \del_\xi \tilde T^{\sigma \xi}    \Big) + (\mu \leftrightarrow \nu)   \nn\\
\label{eq:BoxDeltaLambda}
&= e^{-\s_{(4D)}} \theta^{\nu \rho} \Big( \bar \g^{\mu \sigma}_{(\cK)} \del_{\rho} \del_{\sigma} \langle  f \rangle
  + f_{\a \und \b}^{\ \ \und \g}  \theta^{\mu \l} \langle \g^{a \sigma} \rangle k^\a_{\l} \langle \bar J_{\und \g a} \rangle \del_{\r} \del_\sigma \L^{\und \b} \\
& \qquad \qquad \qquad \qquad \qquad \qquad \qquad \qquad \qquad \qquad
 + \theta^{\mu \l} g_{\sigma \l}^{(\cK)} \del_{\r} \del_\xi \tilde T^{\sigma \xi} \Big) 
 + (\mu \leftrightarrow \nu),  \nn \\
\Box_{\bar G_{(4D)}} \d_{c} \g^{\mu \nu}_{(4D)}  &=  \theta^{\mu \l} \theta^{\nu \r}
 \del_{\r} \Box_{\bar G_{(4D)}} c^{A} \langle L_{A \l}\rangle + (\mu \leftrightarrow \nu)  \nn\\  
&= e^{-\bar \s_{(4D)}} \theta^{\m \l} \theta^{\n \r} \left( \langle L^A_{\l}\rangle \theta^{\sigma \xi} \del_{\r}\del_{\sigma} \langle L_{A \xi} f \rangle
  +  \langle L^A_{\l} \rangle \langle L_{A \sigma} \rangle \del_{\r} \del_\xi \tilde T^{\sigma \xi}  \right) + (\mu \leftrightarrow \nu)   \nn\\
\label{eq:BoxDeltaC}
&= e^{-\bar \s_{(4D)}} \theta^{\nu \r} \left( - (\theta \eta \theta)^{\mu \l} \del_{\r} \del_{\l} \langle  f \rangle
  + \theta^{\mu \l} \eta_{\l \sigma} \del_{\r} \del_\xi \tilde T^{\sigma \xi}   \right) + (\mu \leftrightarrow \nu).
\end{align}
To arrive at \eqref{eq:BoxDeltaLambda} we used that $k^\a_a \bar J_{\a b} = \bar g_{ab}^{(\cK)} = \const$ for type A and C. We also used the 
notation $\bar \g_{(\cK)}^{\m \n} = \theta^{\m \l}\theta^{\n \r} \bar g_{\l \r}^{(\cK)}$.
Now the $f_{\a\und \b}^{\ \ \und\g}$ term in \eqref{eq:BoxDeltaLambda} drops out using \eq{J-average}, 
since, the $\l_\a$, $\a \in \{ 4,5,6 \}$ mutually commute.\footnote{Recall the redefinition of the $\l_\a$ performed in Section~\ref{sec:ZeroModes}.} 
Using that $\bar \g_{(4D)}^{\m \n} = - (\theta \eta \theta)^{\mu \nu} + \bar \g_{(\cK)}^{\m \n}$, \cf \eqref{gbar-explicit}, we note that the $\langle f\rangle$ terms from \eqref{eq:deltaGamma}, \eqref{eq:BoxDeltaLambda}, and \eqref{eq:BoxDeltaC} cancel in \eqref{lin-Einstein-general}, we 
obtain the linearized Einstein tensor 
\begin{align}
 \d \cG^{\mu\nu}  &= \tfrac 12 e^{-2 \bar \s_{(4D)}}\theta^{\nu\nu'} \theta^{\mu\mu'} \bar g_{\mu'\d} \del_{\nu'}\del_\g \tilde T^{\d\g}  
+ (\mu\leftrightarrow \nu).
\end{align}
Here we used the notation $\bar g_{\mu \nu}$ for the restriction of the induced background metric $\bar g_{ab}$  to $a,b = \m, \n \in \{ 0, \dots, 3 \}$, \cf \eqref{gbar-explicit}.
In particular, the Einstein tensor due to the zero modes vanishes  wherever the energy-momentum tensor vanishes. 
For the $c^\mu$ modes this generalizes an observation by Rivelles \cite{Rivelles:2002ez}.
We expect that this result is not restricted to the particular compactifications under consideration here, but should apply
for more general compactifications.

To understand better the response to matter, consider a perturbation $T_{\mu\nu}$ localized 
in some compact region. 
This induces a perturbation in $\L^{\a}$ and $c^A$ similar to the electromagnetic potential of a dipole
with strength $\sim T_{\mu\nu}$, which certainly does not produce the appropriate gravitational metric. 
However if  $\cM^4 \subset \R^{10}$ has a non-trivial embedding  such that 
$\nabla\langle \bar J_{\a}\rangle \neq 0$ or $\nabla\langle L^A\rangle \neq 0$, some
non-derivative coupling to $T_{\mu\nu}$ would arise, leading to gravity-like perturbations of the Ricci tensor,
cf.\ \cite{Steinacker:2009mp}. Such backgrounds arise naturally 
e.g.\ for  cosmological solution \cite{Klammer:2009ku}
or  in the mass--deformed matrix model where $\cM ^4\subset dS^9 \subset \R^{10}$.
More generally, a non-trivial background of massless modes should also lead to such an effect, 
analogous to \cite{Steinacker:2012ra}. 
These issues must be studied in more detail elsewhere. In any case, 
Ricci-flat metric perturbations in vacuum as found above are certainly an essential and encouraging
ingredient, which support the idea to obtain gravity on branes in the uncompactified matrix model.

\section{Conclusions and outlook}

We have studied in detail new brane solutions of the IKKT model with geometry $\cM^4 \times \cK \subset \R^{10}$,
with compact extra dimensions stabilized by angular momentum.  
It turns out that $\cK$ and its moduli contribute to the effective 4-dimensional metric, 
mediated by the non-commutative structure of the brane. 
We focused on the massless modes originating from global symmetries of the model.
Our main result is that the metric contributions due to these 
zero modes lead to Ricci flat curvature perturbations in vacuum, consistent with 
the picture of emergent gravity on the brane. This result is expected to be quite generic,
independent of the specifics of the compact space $\cK$.
On the other hand, the non-derivative coupling to 
the energy-momentum tensor required for gravity -- which can arise in the 
presence of extrinsic curvature \cite{Steinacker:2009mp,Steinacker:2012ra,Steinacker:2012ct} -- 
turns out to cancel for the backgrounds under consideration. 
The reason seems to be that the radial moduli are  stabilized by the non-vanishing flux on $\cK$.
This suggests that other, less rigid types of backgrounds should be considered
in oder to obtain physical gravity  on the brane; this will be 
pursued elsewhere.

The present solutions are also of interest as building blocks for reducible, block-diagonal solutions
of the matrix model,
which may lead to gauge theories on the brane with non-simple gauge groups. 
In suitable configurations, this might allow to obtain (extensions of) the standard model,
cf. \cite{Chatzistavrakidis:2011gs,Aoki:2010gv}. In particular, it would be interesting to study fermions 
on such backgrounds, and to determine whether chiral zero modes arise due to the presence of 
flux on $\cK$. Such chiral zero modes do not arise for static
compactifications e.g. with fuzzy spheres \cite{Chatzistavrakidis:2009ix},
but they might arise here due to the extra rotation of the new solutions.
Moreover, bound states of similar compactifications may enlarge the class of solutions,  
and stabilize them if required. All these are topics for further work.

\paragraph{Acknowledgments.}

H.S. would like to thank the high energy group of CUNY for hospitality and support for a visiting position,
where this work was initiated.
The work of A.P. is
supported by National Science Foundation under grant PHY-1213380 and by a PSC-CUNY grant.
The work of H.S. and J.Z. is supported by the Austrian Fonds f\"ur Wissenschaft und Forschung under grant P24713.

\appendix

 \section{Conserved currents}
\label{sec:sec:currents}

\subsection{The general setup}

We first want to discuss the determination of currents in an abstract setting, following \cite{JochenDiplom}. 
The matrix model is characterized by a set of variables $\Conf = \{ Z_i \in \Alg \}_{i\in I}$ taking values in the $*$-algebra 
$\Alg = \Mat(N \times N, \C)$ and a map $L: \Conf \to \B$, called the \emph{Lagrangian}. Here, $\B$ denotes the 
subset of hermitean elements of $\Alg$. We assume that the Lagrangian is a polynomial in the $Z_i$. 
The elements $Z_i$ may carry a supplementary Grassmann coordinate. The \emph{action} is given by the trace over the Lagrangian. 
In the particular case of the IKKT model, $\{ Z_i \} = \{ X^A, \Psi_\alpha \}$. An infinitesimal variation of $L$ by 
$Z_i \to Z_i + \delta Z_i$ can be written as
\begin{align*}
  \delta L = \sum_{i a} \delta L^{(1)}_{ia} \delta Z_i \delta L^{(2)}_{ia}.
\end{align*}
This leads to the equations of motion
\begin{align}
\label{eq:eom_general}
 \sum_a (-1)^{\pi_{ia}} \delta L^{(2)}_{ia} \delta L^{(1)}_{ia} = 0.
\end{align}
Here $\pi_{ia}$ is determined by the Grassmann parity of $\delta L^{(k)}_{ia}$ and $\delta Z_i$.

Let us now discuss symmetries. We employ the following definition:
\begin{definition}
A \emph{continuous symmetry} consist of maps 
\begin{align*}
 & \alpha_i: \R \times \Conf  \to \Alg,
 & \beta: \R \times \Alg \to \Alg, 
\end{align*} 
which are differentiable in the first variable and fulfill
\begin{align*}
 \alpha_i(0)(Z) & = Z_i,  & \beta(0)(A) & = A,
\end{align*}
and 
\[
 \beta(t) \left( L[\alpha_i(t)(Z)] \right) = L[Z_i]
\]
for all $\{ Z_i \} \in \Conf$ and all $t$. Furthermore, we require that $\beta(t)$ is a $*$-homomorphism.
\end{definition}
This definition is a reflection of the fact that in ordinary field theory, one requires that the action of the symmetry on the Lagrangian can be absorbed in an action on the Lagrangian. For example, a translation of the fields $\phi$ can be absorbed in an opposite translation of $L[\phi]$.

Differentiation \wrt $t$ at $t = 0$ yields
\begin{align}
 \dot \beta L[Z] + \sum_{ia} \delta L^{(1)}_{ia} \dot \alpha_i(Z) \delta L^{(2)}_{ia} = 0.
\end{align}
Using the equation of motion \eqref{eq:eom_general}, we obtain that
\begin{equation}
\label{eq:ConservationLaw}
 \dot \beta L[Z] + \frac{1}{2} \sum_{ia} \left( \{ \delta L^{(1)}_{ia}, \dot \alpha_i(Z) \delta L^{(2)}_{ia} \}_\pm + \{ \delta L^{(1)}_{ia} \dot \alpha_i(Z), \delta L^{(2)}_{ia} \}_\pm \right) = 0
\end{equation}
on-shell. Here $\{ \cdot, \cdot \}_\pm$ denotes the commutator or anti-commutator, depending on the Grassmann parity. Here we chose a symmetric way of commuting the $\delta L^{(k)}_{ia}$.\footnote{Choosing a different commutation procedure 
leads to a different form of conservation law. This corresponds to the usual ambiguity in the definition of a current.
} This is the \emph{conservation law} corresponding to the symmetry. In a semi-classical limit, where (anti-) commutators are replaced by the Poisson bracket, we may use $\theta^{ab} \del_a A \del_b B = \sqrt{|\theta|} \del_a( \sqrt{|\theta|^{-1}} \theta^{ab} A \del_b B)$ to obtain a conserved current.

In order to determine the conservation laws for arbitrary symmetries, we give the $\delta L^{(k)}_{ia}$ for the IKKT model, i.e., for the Lagrangian \eqref{S-YM}:
\begin{align*}
 \delta L^{(1)}_{A 1} & = \tfrac{1}{2}, & \delta L^{(2)}_{A 1} & = X^B [X_A, X_B], \\
 \delta L^{(1)}_{A 2} & = - \tfrac{1}{2} X^B, & \delta L^{(2)}_{A 2} & = [X_A, X_B], \\
 \delta L^{(1)}_{A 3} & = \tfrac{1}{2} [X_A, X_B], & \delta L^{(2)}_{A 3} & = X^B, \\
 \delta L^{(1)}_{A 4} & = - \tfrac{1}{2} [X_A, X_B] X^B, & \delta L^{(2)}_{A 4} & = 1, \\
 \delta L^{(1)}_{A 5} & = \tfrac{1}{2} \tilde \gamma_A^{\alpha \beta} \psi_\alpha, & \delta L^{(2)}_{A 5} & = \psi_\beta, \\
 \delta L^{(1)}_{A 6} & = - \tfrac{1}{2} \tilde \gamma_A^{\alpha \beta} \psi_\alpha \psi_\beta, & \delta L^{(2)}_{A 6} & = 1, \\
 \delta L^{(1)}_{\alpha 1} & = \tfrac{1}{2}, & \delta L^{(2)}_{\alpha 1} & = \tilde \gamma_A^{\alpha \beta} [X^A, \psi_\beta], \\
 \delta L^{(1)}_{\alpha 2} & = - \tfrac{1}{2} \tilde \gamma_A^{\alpha \beta} \psi_\beta, & \delta L^{(2)}_{\alpha 2} & = X^A, \\
 \delta L^{(1)}_{\alpha 3} & = \tfrac{1}{2} \tilde \gamma_A^{\alpha \beta} \psi_\beta X^A, & \delta L^{(2)}_{\alpha 3} & = 1.
\end{align*}

\subsection{The Lorentz current}
\label{app:LorentzCurrent}

The IKKT action \eqref{S-YM} is invariant under the Lorentz symmetry
\begin{align*}
 \delta \Psi_\alpha & = \tilde \lambda_\alpha^\beta \Psi_\beta, & \delta X^A & = \lambda^A_B X^B,
\end{align*}
where $\tilde \lambda$ is a generator of the connected component $\Spin_0(9,1)$ of the $\Spin(9,1)$ group, and $\lambda$ the corresponding generator of the connected component $SO_0(9,1)$ of the Lorentz group. This follows from the $\gamma$ matrix transformation law
\begin{equation}
\label{eq:gammaTransformation}
 \tilde \lambda^\alpha_\beta \tilde \gamma_A^{\beta \gamma} + \tilde \gamma_A^{\alpha \beta} \tilde \lambda^\gamma_\beta + \tilde \gamma_B^{\alpha \gamma} \lambda_A^B = 0.
\end{equation}
This symmetry is internal in the sense that $\dot \beta = 0$. For the corresponding conservation law, we obtain
\begin{align*}
 0 & = \tfrac{1}{2} \lambda^A_C [X_B, \{ X^C, [X_A, X^B] \} ] \\
 & - \tfrac{1}{4} \lambda^A_C \left( \tilde \gamma_A^{\alpha \beta} \{ \Psi_\alpha, X^C \Psi_\beta \} + \tilde \gamma_A^{\alpha \beta} \{ \Psi_\alpha X^C, \Psi_\beta \} - \tilde \gamma_A^{\alpha \beta} [ \Psi_\alpha \Psi_\beta, X^C ] \right) \\
 & + \tfrac{1}{4} \tilde \lambda^\gamma_\alpha \tilde \gamma_A^{\alpha \beta} \left( - \{ \Psi_\gamma, [X^A, \Psi_\beta] \} + \{ \Psi_\beta, \Psi_\gamma X^A \} + [ \Psi_\beta \Psi_\gamma, X^A] -  \{ \Psi_\beta X^A, \Psi_\gamma\} \right). 
\end{align*}
Here we already simplified the bosonic part. Replacing (anti-) commutators by Poisson brackets, we obtain the following conservation law in the semi-classical limit:
\[
 \del_{a} \left[ \sqrt{|\theta|^{-1}} \theta^{ab} \left( \lambda^A_C g_{bd} x^C \theta^{cd} \del_c x_A - \tfrac{i}{2} \lambda^A_C \tilde \gamma^{\a \b}_A X^C \psi_\a \del_b \psi_\b + \tfrac{i}{2} \tilde \lambda_\a^\c \tilde \gamma^{\a \b}_A \psi_{\b} \psi_{\c} \del_b x^A \right) \right] = 0.
\]
Dropping the fermionic part, the semi-classical conservation law is hence
\[
 0 = \del_a ( \sqrt{|\theta|^{-1}} \gamma^{ab} \lambda^A_C x^C \del_b x_A) = \del_a (\sqrt{-|G|} G^{ab} J_b)
\]
with
\begin{equation}
\label{current-general}
 J_b = \lambda^A_C x^C \del_b x^A.
\end{equation}

\subsection{The energy-momentum tensor}

Using our procedure, one may also compute the energy-momentum tensor. For an arbitrary hermitean $\eps$, we have the symmetry $\delta_\eps Z_i = i [\eps, Z_i]$. It leads to $\delta_\eps L = i [\eps, L]$, so this is a symmetry in our sense. For the conservation law \eqref{eq:ConservationLaw}, we compute,
\begin{multline}
\label{eq:EnergyConservation}
 0 = - \tfrac{i}{2} [X_B, \{ [\eps, X_A], [ X^A, X^B ] \} ] - \tfrac{i}{2} \tilde \gamma_A^{\alpha \beta} \left( \{ \Psi_\alpha, [X^A, \eps] \Psi_\beta \} + [X^A, \Psi_\alpha [\Psi_\beta, \eps]] \right)\\
 - \tfrac{i}{4} [\eps, [X^A, X^B][X_A, X_B] + 2 \tilde \gamma_A^{\alpha \beta} \Psi_\alpha [X^A, \Psi_\beta] ]
\end{multline}
For $\eps = X^C$, we can write this as
\begin{equation}
\label{eq:EnergyConservation_2}
 [X_B, T^{BC}] = - \tfrac{i}{2} \tilde \gamma_A^{\alpha \beta} \{ \Psi_\alpha, [X^A, X^C] \Psi_\beta \}
\end{equation}
with
\begin{multline*}
 T^{BC} = \tfrac{i}{2} \{ [X^C, X_A], [ X^A, X^B ] \} + \tfrac{i}{2} \eta^{BA} \tilde \gamma_A^{\alpha \beta} \Psi_\alpha [\Psi_\beta, X^C]  \\
 + \tfrac{i}{4} \eta^{BC} \left( [X^A, X^B][X_A, X_B] + 2 \tilde \gamma_A^{\alpha \beta} \Psi_\alpha [X^A, \Psi_\beta] \right).
\end{multline*}
In the semi-classical limit, the \rhs of \eqref{eq:EnergyConservation_2} vanishes, and we obtain the usual conservation law.

\section{Explicit examples.}
\label{sec:app-explicit}

\subsection{Type A solutions.}
\label{app:Explicit_A}

The generators $\l_\a \in \mso(6)$ in \eqref{eq:Ansatz}, \eqref{eq:U_lambda}
may be chosen as
\begin{align}
\l_4^{(A)} = \begin{pmatrix}
          0 & \one_2 & 0 \\
          -\one_2 & 0 & 0 \\
          0 & 0 & 0
         \end{pmatrix} ,  \quad
\l_5^{(A)} = \begin{pmatrix}
          i & 0 & 0 \\
          0 &  i  & 0 \\
          0 & 0 &  0
         \end{pmatrix} 
\end{align}
along with
\begin{align}
 \l_6^{(A)} &= \begin{pmatrix}
          i & 0 & 0 \\
          0 & 0 & {\scriptsize{\begin{pmatrix}
                   1 & 0 \\ 0 & 0
                  \end{pmatrix}}}  \\
          0 & {\scriptsize{\begin{pmatrix}
                   -1 & 0 \\ 0 & 0
                  \end{pmatrix}}}  & 0 
         \end{pmatrix} , \quad 
 \l_7^{(A)} = \begin{pmatrix}
          0  & 0 & 0 \\
          0 & 0 & {\scriptsize{\begin{pmatrix}
                   0 & 0 \\ 0 & 1
                  \end{pmatrix}}} \\
          0 &  {\scriptsize{\begin{pmatrix}
                   0 & 0 \\ 0 & -1
                  \end{pmatrix}}} & 0 
         \end{pmatrix}, \quad 
\l_8^{(A)} = \begin{pmatrix}
          i & 0 & 0 \\
          0 &  0  & 0 \\
          0 & 0 & 0
         \end{pmatrix}  
\end{align}
where $i \cong {\scriptsize{\begin{pmatrix}
                0 & 1 \\ -1 & 0
               \end{pmatrix}}}$ in complex notation. 
These are clearly two commuting sets of matrices and satisfy
$-\l_4^2 + \l_5^2 = 0$. We use $z_0 = (1,0,0,0,0,0)$ and obtain
\[
 g_{\a \b}^{(S^5)} =
\begin{pmatrix} 
1 & 0 & 0 & 0 & 0 \\
0 & 1 & \cos^2 \varphi^4 & 0 & \cos^2 \varphi^4 \\
0 & \cos^2 \varphi^4 & \cos^2 \varphi^4 + \cos^2 \varphi^5 \sin^2 \varphi^4 & 0 & \cos^2 \varphi^4 \\
0 & 0 & 0 & \sin^2 \varphi^4 \sin^2 \varphi^5 & 0 \\
0 & \cos^2 \varphi^4 & \cos^2 \varphi^4 & 0 & \cos^2 \varphi^4
\end{pmatrix}.
\]

Let us give explicit examples of solutions of \eqref{eq:eom_2}. For simplicity, we choose the standard 
symplectic form  \eqref{Poisson-full} with $\theta^{01} = \theta^{23} = \xi$. 
Furthermore, we assume $r=1$ and $p=1$ in \eqref{k-orthog-A-general}. A solution with $\Theta \neq 0$ is then given by
\begin{align*}
 k^{4} & = (1, \sqrt{3}, 0, 0,1,0), & k^{5} & = (0, 0, 0, 0,0,1).
\end{align*}
Straightforward calculations show that the induced and effective metrics are constant, and 
the effective 4-dimensional metric $\gamma^{\mu\nu}_{(4D)}$ \eqref{eff-metric-4D} has Minkowski signature.
One also  checks that
\begin{equation}
\label{eq:eta4_inequality}
\eta_{\mu\nu} \theta^{\mu\mu'}\theta^{\nu\nu'} g_{\mu\nu}^{(\cK)} < 0,
\end{equation}
which is important for the stabilization of the radius at a nonzero value, 
cf.\ Section~\ref{sec:FluxStabilization}. The generator $\tilde \lambda^6$ introduced in Section~\ref{sec:ZeroModes} can be chosen as $\tilde \l_6 = \diag(0,0,i)$.

\subsection{Type B solutions.}
\label{app:Explicit_B}

For the generators, we may choose
\begin{align}
\l_4^{(B)} = 5\begin{pmatrix}
          0 & \one_2 & 0 \\
          -\one_2 & 0 & 0 \\
          0 & 0 & i 
         \end{pmatrix} ,  \quad
\l_5^{(B)} =4 \begin{pmatrix}
          i & 0 & 0 \\
          0 &  i  & 0 \\
          0 & 0 &  -i
         \end{pmatrix}
\end{align}
along with
\begin{align}
 \l_6^{(B)} &= 3\begin{pmatrix}
          i & 0 & 0 \\
          0 & 0 & \one_2 \\
          0 & -\one_2 & 0 
         \end{pmatrix},  \quad 
 \l_7^{(B)} = \begin{pmatrix}
          0  & 0 & 0 \\
          0 & 0 & {\scriptsize{\begin{pmatrix}
                   1 & 0 \\ 0 & -1
                  \end{pmatrix}}} \\
          0 &  {\scriptsize{\begin{pmatrix}
                   -1 & 0 \\ 0 & 1
                  \end{pmatrix}}} & 0 
         \end{pmatrix}, \quad 
\l_8^{(B)} = \begin{pmatrix}
          i & 0 & 0 \\
          0 &  0  & 0 \\
          0 & 0 & 0
         \end{pmatrix}  
\end{align}
which are appropriately commuting and  satisfy
$-\l_4^2 + \l_5^2 + \l_6^2 = 0$. With $z_0 = (1, 0, 0, 0, 0, 0)$, we obtain
\begin{equation}
\label{kappa-explicit-B}
 g_{\a \b}^{(S^5)} =
\begin{pmatrix} 
25 & 0 & 0 & 0 & 0 \\
0 & 16 & 12 \cos^2 5 \varphi^4 & 0 & 4 \cos^2 5 \varphi^4 \\
0 & 12 \cos^2 5 \varphi^4 & 9 & 3 \cos 8 \varphi^5 \sin^2 5 \varphi^4 & 3 \cos^2 5 \varphi^4 \\
0 & 0 & 3 \cos 8 \varphi^5 \sin^2 5 \varphi^4 & \sin^2 \varphi^5 & 0 \\
0 & 4 \cos^2 5 \varphi^4 & 3 \cos^2 5 \varphi^4 & 0 & \cos^2 5 \varphi^4
\end{pmatrix}.
\end{equation}
For the momenta, we make an ansatz
\begin{align}
k^{4} &= (0,k^{4}_1,0,0,1,0)  \nn\\
k^{5} &= (k^{5}_0,0,0,0,0,1)  \nn\\
k^{6} &= (0,0,k^{6}_2,0,0,0) \ .
\label{type-B-ansatz-k}
\end{align}
Then the condition $\Theta^{\a\b} = 0$ reduces to
\begin{align}
k^{4}_1 k^{5}_0 \theta^{01} &= \xi ,
\label{Theta-vanish-B-ex}
\end{align}
and the different $k^{\a}$ are automatically orthogonal.
The orthogonality condition becomes $\g^{\mu\nu}_{(4D)}k^{(\a)}_\mu k^{(\b)}_\nu + \xi^2 \diag(1,1,0) = p^2 \diag(-1,1,1)$, i.e.
\begin{align}
(k^{4}_1)^2 (\theta^{01})^2  &=  p^2 + \xi^2   \nn\\
(k^{5}_0)^2 (\theta^{01})^2  &=  p^2 -\xi^2   \nn\\
(k^{6}_2)^2 (\theta^{23})^2  &=  p^2  
\label{k-i-equations}
\end{align}
The second together with \eq{Theta-vanish-B-ex}
gives
\begin{align}
(k^{4}_1)^{2} - (k^{5}_0)^{2} &= (k^{4}_1)^{2} -(k^{4}_1)^{-2} \frac{\xi^2}{(\theta^{01})^{2}}
= 2 \frac{\xi^2}{(\theta^{01})^2}  
\label{k-4-equation}
\end{align}
hence 
\begin{align}
 (k^{4}_1)^2 = \frac{\xi^2}{(\theta^{01})^{2}}+\sqrt{\frac{\xi^4}{(\theta^{01})^{4}} + \frac{\xi^2}{(\theta^{01})^{2}}}, 
\label{k-4-equation-2}
\end{align}
and subsequently  $k^{5}_0$ and $p^2$ are determined by \eq{k-i-equations}.
Then the last equation can always be solved for $k^{6}_2$.
The effective 4-dimensional metric is given by
\begin{align}
 \g^{\mu\nu}_{(4D)} 
  = \begin{pmatrix}
    (\theta^{01})^2 + (k^{5}_0)^2 g^{(S^5)}_{55} & 0 & k^{5}_0 k^{6}_2 g^{(S^5)}_{46} & 0 \\
    0 & - (\theta^{01})^2 + (k^{4}_1)^2 g^{(S^5)}_{44} & 0 & 0 \\
    k^{5}_0 k^{6}_2 g^{(S^5)}_{46}(\varphi^4) & 0 & (\theta^{23})^2 + (k^{6}_2)^2 g^{(S^5)}_{66}  & 0 \\
    0 & 0 & 0 & (\theta^{23})^2 
    \end{pmatrix}
\end{align}
This has Minkowski signature provided 
\begin{align}
(\theta^{01})^2 > (k^{4}_1)^2 g^{(S^5)}_{44} 
 \label{mink-signature-type-B}
\end{align}
which is satisfied for suitable parameters $\theta^{01},r,\xi$, in view of \eq{k-4-equation-2}.
Therefore there is indeed a non-empty moduli space 
of type B solutions with the desired Minkowski metric.

\subsection{Type C solutions.}
\label{app:Explicit_C}

Here we may choose the generators $\lambda_\alpha$ as
\begin{align}
\l_4^{(C)} = \begin{pmatrix}
          0 & \one_2 & 0 \\
          -\one_2 & 0 & 0 \\
          0 & 0 & i 
         \end{pmatrix} ,  \quad
\l_5^{(C)} = \begin{pmatrix}
          i & 0 & 0 \\
          0 &  i  & 0 \\
          0 & 0 & 0
         \end{pmatrix}  ,  \quad
\l_6^{(C)} = \begin{pmatrix}
         0 & 0 & 0 \\
          0 & 0  & 0 \\
          0 & 0 & i
       \end{pmatrix}, 
\end{align}
and
\begin{align}
 \l_7^{(C)} = \begin{pmatrix}
          0  & 0 & 0 \\
          0 & 0 & {\scriptsize{\begin{pmatrix}
                   0 & 0 \\ 0 & 1
                  \end{pmatrix}}} \\
          0 &  {\scriptsize{\begin{pmatrix}
                   0 & 0 \\ 0 & -1
                  \end{pmatrix}}} & 0 
         \end{pmatrix}, \quad 
\l_8^{(C)} = \begin{pmatrix}
          i & 0 & 0 \\
          0 &  0  & 0 \\
          0 & 0 & 0
         \end{pmatrix}.
\end{align}
Here $\l_4, \l_5, \l_6$ resp.\ $\l_7,\l_8$ are mutually commuting. We choose $z_0 = (1,0,0,0,1,0) / \sqrt{2}$ and compute
{\small
\[
 g_{\a \b}^{(S^5)} =
\begin{pmatrix} 
2 & 0 & 1 & -\sin \varphi^5 \sin \varphi^6 & 0 \\
0 & 1 & 0 & -\cos \vp^5 \sin \vp^4 \sin \vp^{46} & \cos^2 \varphi^4 \\
1 & 0 & 1 & \cos \vp^{46} \sin \vp^4 \sin \vp^5 & 0 \\
-\sin \varphi^5 \sin \varphi^6 & -\cos \vp^5 \sin \vp^4 \sin \vp^{46} & \cos \vp^{46} \sin \vp^4 \sin \vp^5 & Y & 0 \\
0 & \cos^2 \varphi^4 & 0 & 0 & \cos^2 \varphi^4
\end{pmatrix}
\]
}
with
\begin{align*}
 \vp^{46} & = \vp^4 + \vp^6, \\
 Y & = \sin^2 \vp^4 (\cos^2 \vp^6 + \sin^2 \vp^5) + 2 \cos \vp^4 \cos \vp^6 \sin \vp^4 \sin \vp^6 + \cos^2 \vp^4 \sin^2 \vp^6.
\end{align*}
We also compute
\begin{align}
 \Pi := \int d \vp^4 d \vp^5 \ \cU_6 \cU_5 \cU_4\, {\bf z_0}^\dagger\, {\bf z_0}\, \cU_4^* \cU_5^* \cU_6^* = \pi^2 \mathrm{diag}(1,1,1,1,2,2)
\label{Pi-def}
\end{align}
and note that this commutes with $\lambda_{4,5,6}$.

To find explicit solutions, we choose, as for type A, $\theta^{01} = \theta^{23} = \xi$, $r=1$ and $p=1$ in \eqref{k-orthog-C-general}. 
A solution with $\Theta \neq 0$ is then given by
\begin{align*}
 k^{4} & = \left( 1, - \tfrac{\sqrt{4+ t_+ + t_-}}{\sqrt{3}}, 0, 0, 1, 0 \right), \\
 k^{5} & = \left(\tfrac{\sqrt{2/3} (8+ 2 t_- + 2 t_+)^{3/2}}{3} - \tfrac{5(8 + 2 t_- + 2 t_+)^{5/2}}{36 \sqrt{6}} - \tfrac{2 (6 (8 + 2 t_- + 2 t_+))^{1/2}}{3}, \tfrac{6 - 4 t_- + t_-^2 - 4 t_+ + t_+^2}{9}, 0, 0, 0, 1 \right), \\
 k^{6} & = (0, 0, 0, -1, 0, 0),
\end{align*}
with
\[
 t_\pm = \left( \frac{29 \pm 3 \sqrt{93}}{2} \right)^{1/3}.
\]
As for type A, one checks that the induced and the effective metrics are constant, the 
effective 4-dimensional metric \eqref{eff-metric-4D} has Minkowski signature, and 
also \eqref{eq:eta4_inequality} is fulfilled. Of course this is just one arbitrary point of the non-trivial moduli space of 
solutions.

The above sets of $\la_\a$ are of course not unique. 
The reason for using 5 generators is that this allows to parametrize 
the most general perturbations around these backgrounds in terms of the $\cE^\a$. 
This should allow to systematically study the geometric perturbations and their coupling to matter.

\section{Equations of motion at the operator level}
\label{sec:exact-solutions}

For simplicity, let us consider type A. The other types can be treated in complete analogy. As in Section~\ref{sec:FuzzyTorus}, we use a complex notation for the directions in which $\cK$ is embedded. After a change of coordinates, i.e., rotating the coordinate system by $\cO$, \cf \eqref{eq:U_45}, we have two complex matrices $Z^i$, with
\[
 Z_i = c_i e^{i n_5^i k^5_\mu \bar X^\mu} e^{i n_4^i k^4_\mu \bar X^\mu} V_5^{n_5^i} V_4^{n_4^i}, 
\]
where $V_4, V_5$ are the fuzzy torus generators. The $c_i$ are complex numbers corresponding to the first two components of $r \cO^{-1} z_0$ 
in the notation \eqref{eq:U_45}. They fulfill $|c_1|^2 + |c_2|^2 = r^2$. We have
\begin{equation}
\label{eq:dAlembertianCompact}
 \sum_{j=4}^7 [X^j, [X^j, Y]] = \frac{1}{2} \sum_{i=1}^2\left(  [Z_i, [ Z_i^*, Y]] + [Z_i^*, [ Z_i, Y]] \right) = 2 r^2 Y - \sum_{i=1}^2\left(  Z_i Y Z_i^* + Z_i^* Y Z_i \right),
\end{equation}
where we used that the $V_4, V_5$ are unitary.
From the commutation relations
\begin{align*}
 [\bar X^\mu, \bar X^\nu] & = i \theta^{\mu \nu}, & [\bar X^\mu, V_4] & = 0, & [\bar X^\mu, V_5] & = 0, & V_4 V_5 & = q V_5 V_4,
\end{align*}
one immediately concludes that
\begin{align*}
 \eta_{\lambda \rho} [\bar X^\lambda, [\bar X^\rho, \bar X^\mu]] & = 0, & \delta_{jk} [\bar X^j, [\bar X^k, \bar X^\mu]] & = 0,
\end{align*}
so that the equation of motion \eqref{eom-IKKT} for the non-compact directions is fulfilled. It remains to treat the remaining directions.

Let us first look at the d'Alembertian corresponding to the non-compact directions. We obtain
\[
 \eta_{\mu \nu} [\bar X^\mu, [\bar X^\nu, Z^i]] = Z^i (n_4^i k^4_\mu + n_5^i k^5_\mu) \theta^{\mu \lambda} \eta_{\lambda \rho} \theta^{\rho \nu} (n_4^i k^4_\nu + n_5^i k^5_\nu).
\]
For the d'Alembertian corresponding to the compact directions, we find, using \eqref{eq:dAlembertianCompact},
\[
 \sum_{l=4}^7 [X^l, [X^l, Z^i]] = 4 Z^i \sum_{j=1}^2 |c_j|^2 \sin^2 \tfrac{1}{2} \left( (n_4^j k^4_\mu + n_5^j k^5_\mu) \theta^{\mu \nu} (n_4^i k^4_\nu + n_5^i k^5_\nu) - \tfrac{2\pi}{N}(n_4^j n_5^i - n_5^j n_4^i) \right),
\]
so that the equation of motion is solved if, for both $i$,
\begin{multline}
\label{eq:eom_matrix}
 (n_4^i k^4_\mu + n_5^i k^5_\mu) \theta^{\mu \lambda} \eta_{\lambda \rho} \theta^{\rho \nu} (n_4^i k^4_\nu + n_5^i k^5_\nu) \\
+ 4 \sum_{j=1}^2 |c_j|^2 \sin^2 \tfrac{1}{2} \left( (n_4^j k^4_\mu + n_5^j k^5_\mu) \theta^{\mu \nu} (n_4^i k^4_\nu + n_5^i k^5_\nu) - \tfrac{2\pi}{N}(n_4^j n_5^i - n_5^j n_4^i) \right) = 0.
\end{multline}
Having fixed the $n$'s and the $c$'s (which corresponds to fixing $\l_4, \l_5$, $z_0$), we thus have two equations and 8 free parameters (two 4-vectors). Hence, there will in general be many solutions.

Let us investigate in more detail the relation between the solutions of \eqref{eq:eom_matrix} and the semiclassical solutions discussed in Section~\ref{sec:compact-split}. First of all, we note that
\[
 \k_{\a \b} = \sum_{j=1}^2 |c_j|^2 n_\a^j n_\b^j
\]
corresponds to $g_{\a \b}$, \cf \eqref{eq:gS5_AB}. To have a unified notional, we also write
\begin{align*}
 k^4 & = (k^4_\mu, 1, 0), & k^5 & = (k^5_\mu, 0, 1).
\end{align*}
We may then rewrite \eqref{eq:eom_matrix} as
\[
n_\a^i k^\a_\mu \theta^{\mu \lambda} \eta_{\lambda \rho} \theta^{\rho \nu} n_\b^i k^\b_\nu + 4 \sum_{j=1}^2 |c_j|^2 \sin^2 \tfrac{1}{2} \left( n_\a^j k^\a_a \theta^{ab} n_\b^i k^\b_b \right) = 0.
\]
Assuming that the argument of $\sin^2$ is small, we expand, obtaining
\[
 n_\a^i n_\b^i \left( \theta^{\mu \mu'} \theta^{\nu \nu'} \eta_{\mu' \nu'} \theta^{\rho \nu} k^\a_\mu k^\b_\nu - k^\a_a \theta^{aa'} k_{a'}^{\a'} \k_{\a' \b'} k_{b'}^{\b'} \theta^{b' b} k^\b_b \right) \simeq 0.
\]
Using the identification of $\k_{\a \b}$ with $g_{\a \b}$ and $k^\a_a \theta^{ab} k^\b_b = \Theta^{\a \b}$, we see by comparison with \eqref{orth-full} that, in our limit, a semiclassical solution corresponds to a matrix model solution. It is also clear that our limit corresponds to the limit where $\Theta$ is small.


%


\end{document}